\documentclass[12pt,preprint]{aastex}

\usepackage{psfig,rotate}

\shortauthors{Klose et al.}
\shorttitle{The afterglow of GRB 030226}

\usepackage{graphics,latexsym,psfig}
\usepackage{graphicx,rotate}

\def\farcs{\hbox{$.\!\!^{\prime\prime}$}}  
\def\asec{\ifmmode ^{\prime\prime}\else$^{\prime\prime}$\fi}
\def\degs{\hbox{$.\!\!^\circ$}}  

\epsscale{0.8}

\begin{document}

\title{Probing a GRB progenitor at a redshift of z=2: a comprehensive 
       observing campaign of the 
       afterglow of GRB 030226\thanks{Based on observations collected at the
       European Southern Observatory, La Silla and Paranal, Chile  (ESO
       Programmes 70.D-0523 and 70.D-0531).}}

\author{
S. Klose\altaffilmark{1},         
J. Greiner\altaffilmark{2},       
A. Rau\altaffilmark{2},          
A. A. Henden\altaffilmark{3},      
D.H. Hartmann\altaffilmark{4},   
A. Zeh\altaffilmark{1},            
C. Ries\altaffilmark{5},         
N. Masetti\altaffilmark{6},      
D. Malesani\altaffilmark{7},      
E. Guenther\altaffilmark{1},     
J. Gorosabel\altaffilmark{8,9},   
B. Stecklum\altaffilmark{1},      
L.A. Antonelli\altaffilmark{10}, 
C. Brinkworth\altaffilmark{11},  
J.M. Castro Cer\'on\altaffilmark{8},
A.J. Castro-Tirado\altaffilmark{9},  
S. Covino\altaffilmark{12},         
A. Fruchter\altaffilmark{8},      
J.P.U. Fynbo\altaffilmark{13,14}, 
G. Ghisellini\altaffilmark{12},  
J. Hjorth\altaffilmark{14},      
R. Hudec\altaffilmark{15},   
M. Jel\'{\i}nek\altaffilmark{15},   
L. Kaper\altaffilmark{16},       
C. Kouveliotou\altaffilmark{17},   
K. Lindsay\altaffilmark{4},       
E. Maiorano\altaffilmark{6,19},  
F. Mannucci\altaffilmark{20},    
M. Nysewander\altaffilmark{21},   
E. Palazzi\altaffilmark{6},      
K. Pedersen\altaffilmark{14},    
E. Pian\altaffilmark{6,22},       
D. E. Reichart\altaffilmark{21},  
J. Rhoads\altaffilmark{8},         
E. Rol\altaffilmark{22},          
I. Smail\altaffilmark{23},        
N.R. Tanvir\altaffilmark{24},    
A. de Ugarte Postigo\altaffilmark{9}, 
P.M. Vreeswijk\altaffilmark{25},   
R.A.M.J. Wijers\altaffilmark{16},
E.P.J. van den Heuvel\altaffilmark{16} }

\altaffiltext{1}{ Th\"uringer Landessternwarte Tautenburg, Sternwarte 5, 07778 Tautenburg, Germany} 
\altaffiltext{2}{ Max-Planck-Institut f\"ur extraterrestrische Physik, 85741 Garching, Germany}
\altaffiltext{3}{ U. S. Naval Observatory/Universities Space Research Association, Flagstaff Station, Flagstaff, AZ 86001, USA}
\altaffiltext{4}{ Clemson University, Department of Physics and Astronomy, Clemson, SC 29634-0978, USA} 
\altaffiltext{5}{ Wendelstein-Observatorium, Universit\"atssternwarte, 81679 M\"unchen, Germany }
\altaffiltext{6}{ Istituto di Astrofisica Spaziale e Fisica Cosmica, CNR, Sez. di Bologna, Via Gobetti 101, 40129 Bologna, Italy}
\altaffiltext{7}{ International School for Advanced Studies (SISSA-ISAS), via Beirut 2-4, I-34014 Trieste, Italy} 
\altaffiltext{8}{ Space Telescope Science Institute, 3700 San Martin Drive, Baltimore, MD 21218-2463, USA} 
\altaffiltext{9}{ Instituto de Astrof\'{\i}sica de Andaluc\'{\i}a (IAA-CSIC), Apartado de Correos, 3.004, E-18.080 Granada, Spain}
\altaffiltext{10}{ INAF, Osservatorio Astronomico di Roma, Monteporzio Catone, Italy}
\altaffiltext{11}{ School of Physics and Astronomy, University of Southampton, Highfield, Southampton SO17 1BJ, UK}
\altaffiltext{12}{ INAF, Osservatorio Astronomico di Brera, via E. Bianchi 46, I-23807 Merate (Lc), Italy }
\altaffiltext{13}{ Department of Physics and Astronomy, University of Aarhus, Ny Munkegade, 8000 Aarhus C, Denmark}
\altaffiltext{14}{ Astronomical Observatory, University of Copenhagen, Juliane Maries Vej 30, 2100 Copenhagen, Denmark} 
\altaffiltext{15}{ Astronomical Institute of the Czech Academy of Sciences, 25165 Ondrejov, Czech Rep.}
\altaffiltext{16}{ University of Amsterdam, Kruislaan 403, 1098 SJ Amsterdam, The Netherlands}
\altaffiltext{17}{ NSSTC, SD-50, 320 Sparkman Drive, Huntsville, AL 35805, USA}
\altaffiltext{18}{ Institute of Astronomy, University of Cambridge, Madingley Road, CB3 0HA Cambridge, UK}
\altaffiltext{19}{ Dipartimento di Astronomia, Universit\`a di Bologna, via Ranzani 1, 40129 Bologna, Italy}
\altaffiltext{20}{ IRA/CNR, Sezione di Firenze, Largo E. Fermi 5, I-50125 Florence, Italy}
\altaffiltext{21}{ University of North Carolina, Chapel Hill, NC, USA}
\altaffiltext{22}{ INAF, Osservatorio Astronomico di Trieste, Via Tiepolo 11, 34131 Trieste, Italy}
\altaffiltext{23}{ Institute for Computational Cosmology, Department of Physics, University of Durham, South Road, Durham DH1 3LE, UK}
\altaffiltext{24}{ Department of Physical Sciences, University of Hertfordshire, College Lane, Hatfield Herts, AL10 9AB, UK}
\altaffiltext{25}{ European Southern Observatory, Alonso de C\'ordova 3107, Vitacura, Casilla 19001, Santiago 19, Chile}

\date{Received: 4 June 2004 / Accepted: 14 July 2004}

\begin{abstract}
We report results from a comprehensive follow-up observing campaign of the
afterglow of GRB 030226, including VLT spectroscopy, VLT  polarimetry, and
\it Chandra \rm X-ray observations. In addition, we present BOOTES-1 wide-field
observations at the time of  the occurrence of the burst. First observations
at ESO started 0.2 days after the event when the  GRB afterglow was at a
magnitude of $R\sim$19 and continued until the afterglow had faded below the
detection threshold ($R>26$). No underlying host galaxy was found. The optical
light curve shows a break around 0.8 days after the burst, which is achromatic
within the observational errors, supporting the view that it was due to a
jetted explosion. Close to the break time the degree of linear polarization of
the afterglow light was less than 1.1\%, which favors a uniform jet model
rather than a structured one. VLT spectra show two absorption line systems at
redshifts $z=1.962\pm0.001$ and at $z=1.986\pm0.001$, placing the lower limit
for the redshift of the GRB close to 2. We emphasize that the kinematics and
the composition of the absorbing clouds responsible for these line systems is
very similar to those observed in the afterglow of GRB 021004. This
corroborates the picture in which at least some GRBs are physically related to
the explosion of a Wolf-Rayet star.
\end{abstract}

\maketitle

\section{Introduction}

Understanding the nature of GRB progenitors remains a primary focus in GRB
research. The spectroscopic identification of supernova (SN) light underlying
the afterglows of the nearby bursts 030329 (Hjorth et al. 2003a; Kawabata et
al. 2003; Matheson et al. 2003; Stanek et al. 2003) and 031203 (Malesani et
al. 2004) convincingly demonstrated that core-collaps supernovae are
physically related to long-duration GRBs. In most cases, however, GRB-SNe are
too faint at their suspected maximum light (cf. Zeh et al. 2004) to confirm
their appearance via spectroscopic observations. Revealing the nature of a GRB
progenitor by indirect methods thus remains an important approach, in addition
to an understanding of the afterglows themselves. In this context of
particular interest is the fine structure detected in two of the best
monitored afterglow light curves (GRB 021004: e.g., Bersier et al. 2003; GRB
030329: e.g., Lipkin et al. 2004) and the kinematics of absorbing clouds
revealed in early-time spectra of the afterglows of GRB 020813 and 021004 (for
a discussion see Chevalier et al. 2004; Mirabal et al. 2003; and Schaefer et
al. 2003), since both features could be signatures of the physical conditions
in the GRB environment.  Here we report the results of our comprehensive
observing campaign of the afterglow of GRB 030226. Even though  the afterglow
light curve started faint and faded rapidly, which did not allow us to monitor
its evolution in great detail, our early-time spectra do reveal strong
spectral similarities between the afterglow of GRB 030226 and those of GRB
020813 and 021004.

GRB 030226 was discovered by the {\it HETE-2} satellite on 2003 February
26.15731 UT (3:46 UT) with a duration in excess of 100 seconds and a peak flux
and fluence in the 30-400 keV band of $\sim$1.2$\times$10$^{-7}$ erg cm$^{-2}$
s$^{-1}$ and 5.7$\times$10$^{-6}$ erg cm$^{-2}$, respectively (Suzuki et
al. 2003). The optical counterpart was detected $\sim$2.5 hours later (Fox et
al. 2003, 2004) at a magnitude $R \sim$ 18.5 (Garnavich et al. 2003) at
coordinates (J2000) R.A. = 11$^{\rm h}$ 33$^{\rm m}$ 04$\fs$92; Decl. =
+25$^{\circ}$ 53$'$ 55$\farcs$6  (Price et al. 2003a; von Braun et al. 2003),
well within the {\it HETE-2}  Wide Field X-ray monitor (WXM) and Soft X-ray
Camera (SXC) error boxes.  No optical emission coincident in time with the GRB
prompt event was detected (Castro-Tirado et al. 2003;  \S~\ref{afterglow}).  A
subsequent {\it Chandra} observation led to the detection of the X-ray
afterglow  at a position consistent with that of the optical transient, thus
confirming its association with GRB 030226 (Pedersen et al. 2004a).

\section{Observations and data reduction \label{data}}

Imaging of the central part of the SXC error circle  in $J_s, H, K_s$ began
$\sim$4.5\,hrs after the burst using VLT/ISAAC at ESO Paranal. Simultaneously,
imaging in the optical in $BVR$ was performed using the ESO NTT telescope at
La Silla equipped with the multi-purpose instrument EMMI.  Also, the 3.8-m
United Kingdom Infra-Red Telescope (UKIRT) at Mauna Kea, Hawaii, started
observing the GRB afterglow 6\,hours after the burst using the UKIRT
Fast-Track-Imager (UFTI). Further imaging was performed during the following
nights with the ESO NTT telescope equipped with the SUperb Seeing Imager
(SuSI2).  Additional data were obtained with the Wendelstein 80-cm telescope
near Munich, Germany (using the direct imaging CCD camera MONICA), the  2.5-m
Nordic Optical Telescope (NOT) equipped with the Andaluc\'{\i}a Faint Object
Spectrograph and Camera ALFOSC, the 1.0-m Jacobus Kapteyn Telescope (JKT), and
the Italian 3.5-m Telescopio Nazionale Galileo TNG, using the focal reducer
instrument DOLORES (all at La Palma, Spain), the Tautenburg 1.34-m Schmidt
telescope equipped with prime focus 4k$\,\times\,$4k CCD camera, the Asiago
1.8-m telescope equipped with the AFOSC camera, UKIRT, and the 8-m  Gemini
South telescope at Cerro Pach\'on, Chile. The data base, and  the derived
magnitudes (not corrected for Galactic extinction; $E(B-V) \approx 0.02$  mag,
Schlegel et al. 1998) are summarized in Table~\ref{log}. These magnitudes
supersede those obtained by this collaboration and published in the GCN
Circular archive.

Regular observations of the sky were also acquired by the wide-field
cameras of the BOOTES project (Castro-Tirado et al. 1999) at the BOOTES-1
astronomical station in Mazag\'on, Spain.  The GRB 030226 location was
serepinditously imaged in a simultaneous, unfiltered frame (180 s exposure
time) taken on 03:45:16 - 03:48:16 UT. Additional images were available prior
to and following the event.

\subsection{Photometry \label{NIRdata}}

After the normal image processing steps (flatfielding, stacking), all stars
were extracted using DAOPHOT as implemented in IRAF\footnote{IRAF is
distributed by the National Optical Astronomical Observatories, which is
operated  by the Associated Universities for Research in  Astronomy, Inc.,
under contract to the National Science Foundation.}. Point-spread-function
(PSF) fitting was used to derive the brightnesses of the afterglow and
comparison stars. PSF stars were carefully selected by first inspecting the
deeper, higher resolution images to ensure that no background contamination
was present and that all objects were stellar. After extraction, ensemble
differential photometry was performed using comparison stars chosen from the
field photometry file provided by Henden (2003). Typically four to six
comparison stars were used, depending on the depth of the image and the
bandpass. No transformations were made to the standard Johnson-Cousins system,
but comparison stars were chosen to reduce transformation effects whenever
possible.

\begin{figure}
\plotone{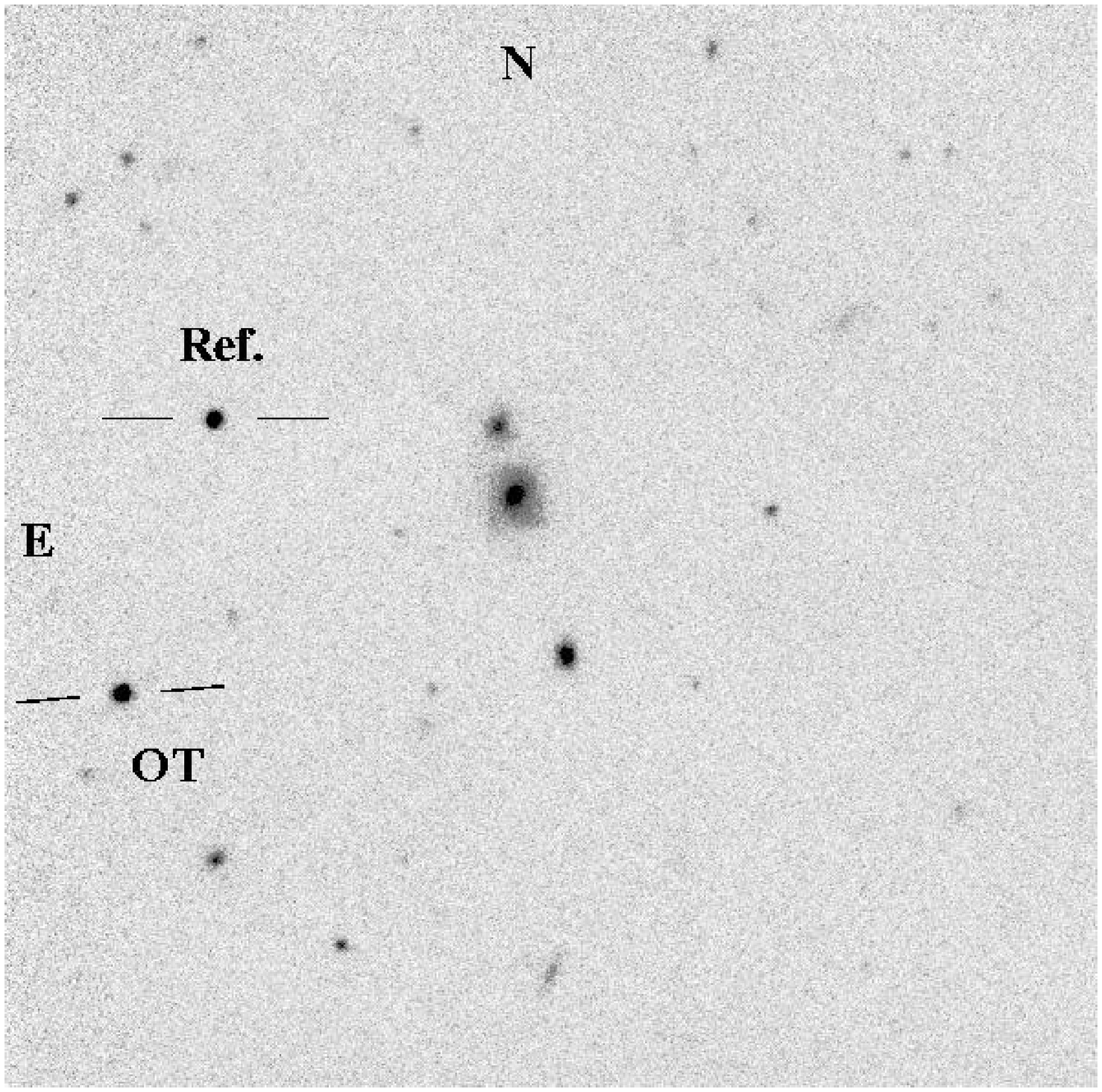}
\caption{VLT/ISAAC $K_s$-band images of the afterglow of GRB 030226 obtained
4.5\,hrs after the burst. The image size is approximately
$100''\,\times\,100''$; `Ref' denotes the reference star, which we used for
the NIR photometry (see \S~\ref{NIRdata}).}
\label{Kimages}
\end{figure}

For the reduction of the near-infrared (NIR) frames ESO's {\it eclipse}
package was used  (Devillard 2002). Photometric calibration of the ISAAC field
was performed by the observation of the UKIRT standard star FS~136 (Hawarden
et al. 2001) with the VLT. Airmass corrections were applied  according to the
coefficients provided on ESO's Web pages (in units of mag airmass$^{-1}$):
$k_J$=0.06, $k_H$=0.06, $k_K$=0.07. An independent calibration of the UKIRT
$K$-band data, made using the standard star FS 130 (Hawarden et al. 2001), is
in good agreement with the one described above.\footnote{Following the
detailed  photometric work performed by Labb\'e et al. (2003) on the \it
Hubble Deep Field South, \rm we assumed that the ISAAC $J_s, H, K_s$  filters
match the faint NIR standard star system (Persson et al. 1998). We also refer
the reader to Labb\'e et al. (2003; their section 2.1)  for  more details on
the $J_s$ and $K_s$ filters.}  Next we deduced the following magnitudes for
the brightest point-like object close to the optical transient (marked ``Ref''
in Fig.~\ref{Kimages}) at R.A. (J2000) =  11$^{\rm h}$ 33$^{\rm m}$ 04$\fs$27,
Decl. (J2000) = 25$^{\circ}$ 54$'$ 22\asec: $J_s=17.10\pm0.10, H=16.54\pm0.10,
K_s=16.44\pm0.05$ \footnote{The optical magnitudes of this star are
$B=20.15\pm0.03, V=19.07\pm0.03, R_c=18.33\pm0.13, I_c=17.92\pm0.18$ (Henden
2003).}. This calibration was confirmed with independent observations using
the Calar Alto 3.5-m telescope equipped with the  Omega Prime NIR camera on
March 6, 2004.

\subsection{VLT spectroscopy}

The early report of the discovery of the GRB afterglow by Fox et al. (2003)
allowed rapid spectroscopic observations of the optical transient,
performed  only 5\,hrs after the burst, on  2003 Feb 26,
08:52--09:07 UT, with VLT Yepun/FORS2. A 900\,s exposure using the 300V grism
was obtained when the afterglow was at a magnitude of $B=19.8$.  Using a
1\farcs0 slit, the 3.3\,\AA/pixel scale leads to a resolution of 13\,\AA \
(FWHM, 870 km/s at $\lambda$=4500 \AA) at 0\farcs9--1\farcs1 seeing.  A
second spectroscopic run was performed one night later, on 2003 Feb 27,
07:35--09:08 UT, using VLT Antu/FORS1 equipped with the 600B grism. Five
exposures, each one lasting 900\,s, were taken at a time  when the afterglow
had a magnitude of $B=21.7$. The 1.2\,\AA/pixel scale gave a resolution of
7\,\AA\ (FWHM, 470 km/s at $\lambda$=4500 \AA) at 1\farcs0--1\farcs2 seeing
with a  1\farcs3 slit. Flatfield and bias correction was applied in the usual
way  using IRAF. Background subtraction was performed and the wavelength
calibration was done using HgCdHe and HgCdHe+Ar calibration lamps. The spectra
were corrected for achromatic slit losses following the procedure described in
Vreeswijk et al. (2004). The standard stars LTT 6248 (spectral type A) and
BD +33\degs2642 (spectral type B2~IV) were observed in order to flux-calibrate
the spectra taken with FORS2 and FORS1, respectively.

\subsection{VLT polarimetry}

Optical polarimetric observations of the afterglow were performed with VLT
Antu/FORS1 and a Bessel $R$ filter. Observations started on 2003 Feb 27.211
UT, approximately 1 day after the GRB, when the afterglow magnitude was about
$R = 20.7$.  Imaging polarimetry is achieved with a Wollaston prism splitting
the image of each object in the field into the two orthogonal polarization
components, which appear in adjacent areas of the CCD image. In this way, the
measurement is insensitive of any variation in the source brightness during
the observation. Two sets of images were acquired with four different angles
(0$^\circ$, 337\fdg5, 315\fdg0, 292\fdg5) and with 15 min exposure each.
Unfortunately, in one exposure a cosmic ray hit the afterglow, making any
procedure unreliable. This set was therefore excluded from the analysis.  A
polarimetric standard star, Hiltner\,652, was also observed in order to fix
the offset between the polarization and the instrumental angles. Image
reduction was performed using the {\it eclipse} tools (Devillard 2002), while
photometry was extracted using the Gaia package\footnote{\texttt
{http://star-www.dur.ac.uk/\~\,$\!$pdraper/gaia/gaia.html}}. The general
procedure followed for FORS1 polarimetric observation analysis is extensively
discussed, e.g., in Covino et al. (1999).

\subsection{{\it Chandra} X-ray observations}

A 40.17 ksec target-of-opportunity {\it Chandra} observation was initiated on
2003 February 27 at 16:49 UT, about 1.54 days after the GRB. Only one chip was
read out (the back-illuminated S3 chip) at a frame time of 0.44 sec. Three
X-ray sources are detected in the 8$\times$8 arcmin$^2$ field of view with a
significance of more than 5$\sigma$, two of which are within the \it
HETE\rm/SXC  error circle. The brightest source is detected with a mean count
rate of 0.010$\pm$0.001 cts s$^{-1}$, at R.A. = 11$^{\rm h}$ 33$^{\rm m}$
04$\fs$9; Decl.  = +25$^{\circ}$ 53$'$ 55\asec\, (J2000), coincident within
the error ($\pm$0\farcs5) with the position of the optical afterglow. This
X-ray source also shows a fading by a factor of three during the observation,
further strengthening its identification as the X-ray afterglow of GRB 030226.
The other two sources are constant within their statistical errors,  and
detected at R.A. = 11$^{\rm h}$ 33$^{\rm m}$ 02$\fs$8; Decl. =  +25$^{\circ}$
53$'$ 55\asec\ and R.A. = 11$^{\rm h}$ 33$^{\rm m}$ 13$\fs$9; Decl. =
+25$^{\circ}$ 52$'$ 29\asec, respectively.

\section{Results \label{results}}

\subsection{The optical/NIR/X-ray light curve \label{alphas} }

\begin{figure}
\plotone{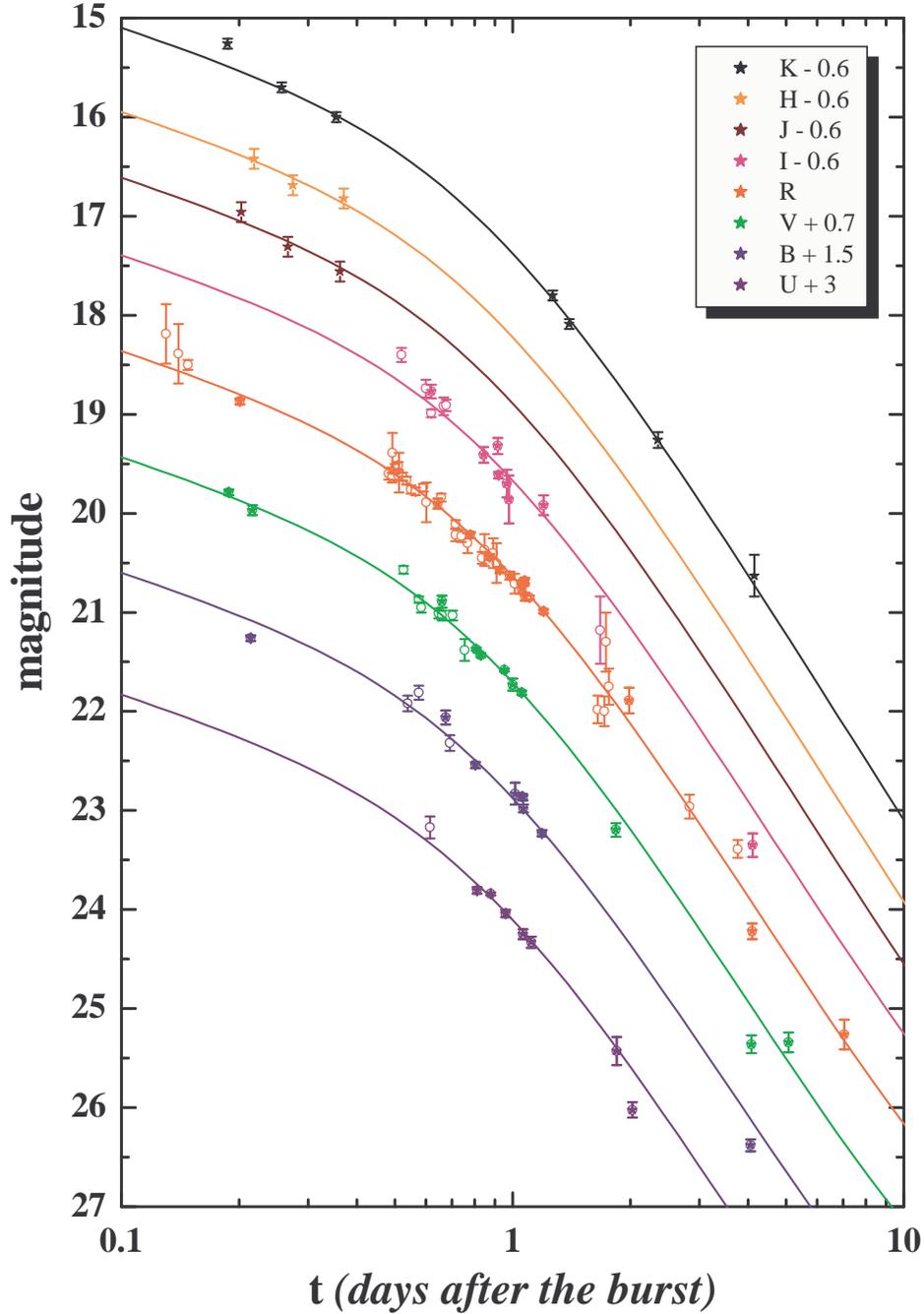}
\caption{Multi-color light curve of the optical transient following GRB 
030226. Data from Table~\ref{log} are shown here (stars) together with data
(open circles) reported in various Gamma-Ray Burst Circulars (Ando et
al. 2003; Garnavich et al. 2003; Price \& Warren 2003; Guarnieri et al. 2003;
von Braun et al.  2003; Rumyantsev et al. 2003a,b; Covino et
al. 2003a; Fatkhullin et al. 2003; Semkov 2003) and by Pandey et
al. (2004). Note that  the $R$-band data reported in GCNs \# 1881, 1882, 1884,
1890, and 1908 were corrected for a 0.4 mag offset at the zero point.}
\label{Rband}
\end{figure}

In order to analyze the light curve of the GRB afterglow, we combined our data
with published data from \it Gamma-Ray Burst Coordinated Network Circulars \rm
(GCNs\footnote{\tt http://gcn.gsfc.nasa.gov/gcn/gcn3\_archive.html\rm};  for
references see Fig.~\ref{Rband}) and Pandey et al. (2004). In doing so, we
followed the standard procedure and fitted a broken power-law decay to all
photometric data.  Of the available multi-wavelength data, the $R$-band light
curve is sampled best. Therefore, we fitted the $R$-band light curve, using
the fitting equation suggested by Beuermann et al. (1999), but in the
representation given by Rhoads \& Fruchter (2001):
\begin{equation}
   F_\nu (t) = 2^{1/n} \ F_{\nu}(t_b)[(t/t_b)^{\alpha_1\,n}+
                    (t/t_b)^{\alpha_2\,n}]^{-1/n} \,.
\label{beuermann}
\end{equation}
Here $F_\nu$ is the flux density\footnote{We use the notation $F_\nu \propto
t^{-\alpha}\, \nu^{-\beta}$ for the time and frequency dependence of the flux
density.}, $\alpha_1$ and $\alpha_2$ are the asymptotic decay indices before
and after the light curve break, $t$ is  the time elapsed since the GRB
trigger, $t_b$ the break time, and $n$ the parameter which describes the
smoothness of the break. In all cases we fitted apparent magnitudes after
correction for Galactic extinction. The best fit of the combined $R$-band data
set is $\alpha_1=0.50\pm0.35$, $\alpha_2 = 2.55\pm 0.38$, $t_b= 0.83\pm0.10$
days, and $n=0.91\pm0.84$  ($\chi^2$/d.o.f.= 2.8; Fig.~\ref{Rband}). While
similar values for $\alpha_2$ have also been found for other afterglow light
curves, the change in decay slope, $\alpha_2-\alpha_1$ is rather large.  
If we use only our $R$-band data (Table~\ref{log}), the best
fit is $\alpha_1=0.70\pm0.25$, $\alpha_2 = 2.66\pm 0.32$, $t_b= 1.04\pm0.12$
days, and $n= 1.2\pm1.1$ ($\chi^2$/d.o.f.= 3.8), consistent with the previous
results.  Since no evidence for an underlying host was found in our data
(\S~\ref{host}), both fits assumed a magnitude for the host of $m=28$ in all
bands. In the following we will use the results of the first fit.

Figure~\ref{OC} shows the difference between observed and best fit magnitudes,
adopting for all bands the light curve parameters  obtained from the $R$-band
fit. Several small-scale fluctuations up to about 0.4 mag in both directions
are apparent. In particular, on day 0.18 the optical transient is notably
brighter in $K_s$ than expected based on the light curve fit. On this day the
afterglow was bright enough to do individual photometry of each NIR frame with
consistent results compared to the stacked frames. This indicates that there
were not any gross flattening errors since the optical transient was dithered
across a fair fraction of the detector, so that some other explanation has to
be invoked as to why this data point does not match the fit properly. Kulkarni
et al. (2003) also noticed an unusual behavior of the afterglow light curve
between 2.3 and 3.3 days after the burst. While flux fluctuations are also
seen in the X-ray light curve  observed with {\it Chandra}, a comparison of
the strongest fluctuations in the $R$ band (seen around 1.7 days after the
GRB) with those in X-rays shows no obvious correlation between them
(Fig.~\ref{xlc}). However, in Fig.~\ref{xlc} four of the five $R$-band  data
points were taken from the literature (GCNs), so that a potential small
mismatch in the photometric calibration cannot be ruled out.

\subsection{The optical spectra \label{cosmology}}

The VLT spectra of the optical transient (Figs.~\ref{fig:forsNight1},
~\ref{fig:forsNight2}) show several absorption lines, but no prominent
emission lines. The latter is not surprising, as the expected lines (e.g.,
[O~II]$\lambda 3727$, [O~III]$\lambda$5007) are located outside our wavelength
band, while Ly\,$\alpha$ in emission is difficult to identify. The absorption
lines are due to rest-frame ultraviolet metallic lines usually present in
spectra of optical afterglows and high redshift galaxies (e.g., Castro et
al. 2003; Mirabal et al. 2003; Masetti et al. 2001, 2003; Savaglio et
al. 2003; Shapley et al. 2003). From these lines two redshift systems at
$z$=1.962$\pm$0.001 and $z$=1.986$\pm$0.001 can be determined, consistent with
earlier reports (Ando et al.  2003; Chornock \& Filippenko 2003; Greiner et
al. 2003a; Price et al. 2003b). In addition to the therein stated absorption
line systems of Fe\,II, Al\,II, and C\,IV, we find systems of O\,I, C\,II,
Si\,IV, and Si\,II.   In our first epoch spectrum we also detect a feature at
5711 and 5726\AA, which may be attributed to redshifted Mg II at $z$=1.042, a
redshift system found via high-resolution spectroscopy with the Keck telescope
(Chornock \& Filippenko 2003; Price et al. 2003b).  Based on our data we
cannot identify the nature of this absorber (see also
Fig.~\ref{hosts}). Whether or not a similar feature at 5494 and 5517\AA \ as
well as 6280 and 6291\AA\ is also due to redshifted Mg II is less certain.

Following the detection of Fe\,II at a rest frame wavelength of 1608\AA \
(oscillator strength $f$=0.058; Prochaska et al. 2001) one would expect to see
also Fe\,II lines at rest frame wavelengths of 2344\AA \ ($f$=0.114), 2586\AA
\  ($f$=0.069), 2600\AA \ ($f$=0.239) and perhaps even 2374\AA \
($f$=0.031). Although the lines are covered by the 300V spectrum of the first
night, no clear line identification can be proposed as they blend with the
telluric features at observer frame wavelengths around 7000 and
7800\AA \ (Fig.~\ref{fig:forsNight1}). 

For the line fitting we used the IRAF/SPLOT task which assumes a
Gaussian profile. Table~\ref{tab:lines} shows the line identifications
together with the observer and rest-frame wavelengths, and the redshifts for
each line detected. No line broadening and thus no velocity dispersion above
the instrumental resolution is observed. In addition, the computed rest frame
equivalent widths (EWs) are listed, which were calculated by dividing the
measured observer frame EWs by $(1+z)$. Based on the measured equivalent
width we calculated the column density, $N$, assuming the optically thin case
(e.g., Spitzer 1968),
\begin{equation}
\label{spitzer}
EW = \frac{e^2}{4\epsilon_0 m_e\,c^2} \ 
      \lambda_{\rm rest}^2\,N\,f\,,
\end{equation}
where the symbols $e, m_e, \epsilon_0$, and $c$ have their usual meaning for
the SI system, and $f$ is the oscillator strength of the corresponding
transition. The resulting column densities are reported in
Table~\ref{tab:lines}. The majority of the observed absorption lines are
strong (EW$_r>$1\,\AA) and may be saturated. In that case they deviate from
the linear part of the curve of growth and the approximation given in
Eq.~(\ref{spitzer}) has to be considered as a lower limit for the derived
column densities.

Finally, the identified lines allow us to refine the redshift of the
GRB to $z$=1.986, or larger. Assuming a flat universe with a matter density
$\Omega_M =$ 0.3, vacuum density (cosmological constant) $\Omega_\Lambda=$
0.7, and Hubble constant $H_0=65$ km s$^{-1}$ Mpc$^{-1}$, this redshift
corresponds to a luminosity distance of $5.12\,\times\,10^{28}$ cm (16.6 Gpc)
and a distance modulus of 46.1 mag. The look-back time is 11 Gyr. The
difference in luminosity distance between both redshift systems is 245 Mpc,
and the difference in velocity is 2410 km s$^{-1}$.

\subsection{The spectral energy distribution of the afterglow \label{SEDsect}}

\begin{figure}
\includegraphics[scale=0.8,angle=0]{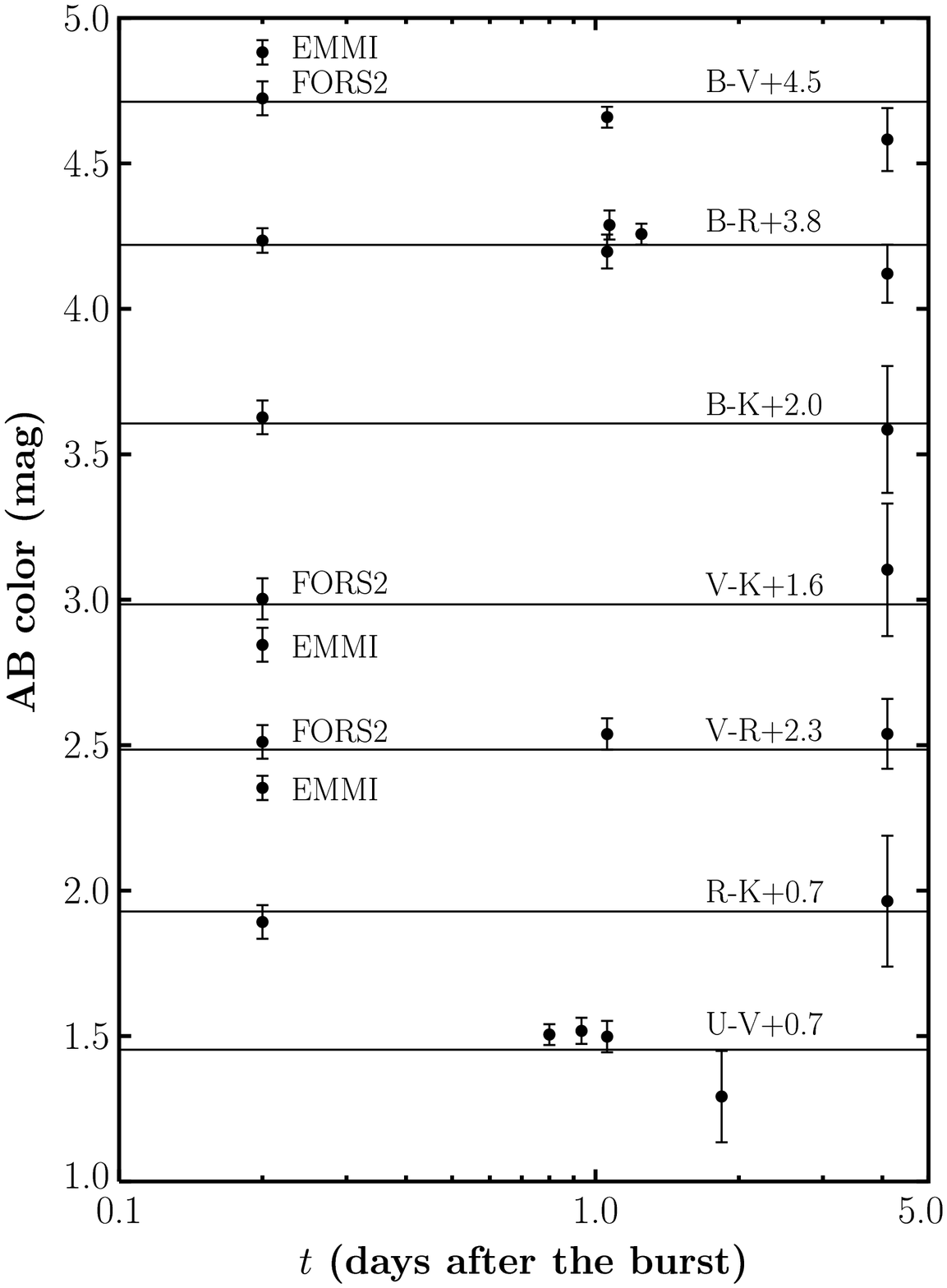}
\caption{The color of the afterglow as a function of time, not
corrected for Galactic extinction. The color was calculated from AB
magnitudes, with the AB offsets given in Table~\ref{ABmags}. Only data from
telescopes with known AB offsets were used. Straight lines give the mean of
the corresponding color. Even though we cannot resolve the origin for the
offset between the NTT/EMMI and VLT/FORS2 $V$-band data during the first
observing epoch, we favor the view that our data (Table~\ref{log}) point to
an achromatic evolution of the afterglow between 0.2 and 4 days after the
burst.}
\label{color}
\end{figure}

Of particular interest is the question whether or not the afterglow exhibited
color evolution, as it has been reported for some GRBs (e.g.,  Bersier et
al. 2003). However, given the relatively sparse data for GRB 030226 based on
observing campaigns with different telescopes at different sites, we can only
check for an obvious long-term trend in the color of the afterglow. Such a
trend is not seen in our data (Fig.~\ref{color}). Within the observational
errors the break in the light curve was achromatic, and so was a potential
flux variation on day 4.1 (Fig.~\ref{Rband}). This supports the view that the
break was due to a collimated explosion (Rhoads 1999; M\'esz\'aros \& Rees
1999). On the other hand, around day 0.2 there is an excess flux in the $K$
band (Fig.~\ref{OC}). Since the multi-color data were not taken
simultaneously, we cannot decide whether this is indeed a color variation or
caused by a flux variation as it has been seen in other well-studied
afterglows (e.g., Lipkin et al. 2004).

Based on the above discussion we conclude that the observed spectral energy
distribution (SED) of the afterglow is best described by assuming no color
variations. The shape of the light curve is then the same in all photometric
bands. Using the magnitude-flux conversion according to Bessel (1979) and
Bessel \& Brett (1988), we find that the observed mean spectral slope,
$\beta$, across the $BVRIJHK$ bands, corrected for Galactic extinction, is
$\beta = 0.70\pm0.03$ (Fig.~\ref{SED}), with no evidence for additional
reddening by dust in the GRB host galaxy ($A_V$(host)=0 mag). This holds for
MilkyWay-like dust as well as for SMC-like dust, for any assumed ratio of
total-to-selective extinction between $R_V$=3 and 5. This result differs from
Pandey et al. (2004) who based their conclusions exlusively on optical  data
obtained on days 0.62 and 1.79 after the burst ($\beta \approx 1$). Pandey  et
al. argue that from the theoretical point of view the afterglow data are best
understood if the intrinsic spectral slope was in fact $\sim$0.55, requiring
that the afterglow light was slightly reddened by dust in the GRB host
galaxy. This spectral slope is indeed not much different from the one we
deduce based on our larger data set, including NIR data, but without the need
for additional reddening in the host galaxy. It is clear that the
determination of $\beta$ and the amount of reddening in a GRB host is very
sensitive to the extent and quality of the data utilized. 
 
Is there evidence for dust in the host based on the X-ray data? The \it
Chandra \rm observation took place after the break in the light curve (for
details, see Pedersen et al. 2004b). The unabsorbed X-ray flux in the 0.3--10
keV band (source frame band of 0.9--30 keV) is 7.3$\times$10$^{-14}$ erg
cm$^{-2}$ s$^{-1}$, corresponding to a mean (isotropic) luminosity of
2.4$\times$10$^{45}$ erg s$^{-1}$ (averaged over the time interval of
1.54--2.04 days after the GRB). A free power-law fit to the X-ray spectrum
with just local foreground absorption yields an energy index across the 0.3 to
10 keV band of $\beta_X$=1.04$\pm$0.20 and a hydrogen column density $N_{\rm
H}$ = (4$\pm$1)$\times$10$^{20}$ cm$^{-2}$ (Fig.~\ref{xsp}). Since the
absorbing column is above the Galactic foreground value, we performed a second
fit with two different absorbing columns, namely one fixed at the Galactic
value ($N_{\rm H}$(Gal) = 1.8$\times$10$^{20}$ cm$^{-2}$; Schlegel et
al. 1998) and the other one at a redshift of the host galaxy of $z=1.986$. The
resulting energy index then is $\beta_X$=1.08$\pm$0.15  and the hydrogen
column density in the host  $N_{\rm H}$(host) = (3.2$\pm$1.5)$\times$10$^{21}$
cm$^{-2}$, corresponding to $A_V$(host) = 2.4 mag (for a Galactic dust-to-gas
ratio). Evidence for so much extinction in the host galaxy is definitly not
seen in the optical/NIR data, a discrepancy also known from other afterglows
(Galama \& Wijers 2001; Hjorth et al. 2003b; Stratta et al. 2004).

Unfortunately, our VLT spectra cannot be used to derive the  hydrogen column
density in an independent way. The unfortunate position of Ly${\alpha}$ at the
short edge of our VLT spectra and the uncertainty in the spectral shape
longwards of the Ly${\alpha}$-edge (Fig.~\ref{fig:forsNight1}) do not allow a
quantitative treatment of the Ly${\alpha}$ absorption and therefore the
hydrogen column density from the optical data. On the other hand, our
polarization data also exclude  a large amount of dust in the GRB host galaxy
along the line of sight.

\subsection{The polarization of the afterglow}

Figure~\ref{QU} shows the position of the afterglow in the Stokes $Q$-$U$-
plane. The polarization level $P$ is related to the Stokes parameters through
$P = \sqrt{Q^2+U^2}$, while the polarization angle is $\vartheta =
\frac{1}{2}\arctan(U/Q)$. Our measurement yields for the afterglow $Q =
-0.0028 \pm 0.0030$, $U = 0.0014 \pm 0.0048$ (1$\sigma$ error), which
translates, after correcting for the polarization bias (Wardle \& Kronberg
1974), to a strict upper limit $P < 1.1\%$ (2$\sigma$). This is one of
the lowest limits ever reported for a GRB afterglow  (e.g., Covino et
al. 1999, 2003b; Rol et al. 2003; Barth et al. 2003), even if values as low as
$P \approx 0.5\%$ have been measured for GRB\,030329 (Greiner et
al. 2003b). The low polarization of the afterglow of GRB 030226 argues against
substantial dust extinction in the GRB host along the line of sight, provided
that the dust properties there are not much different from those of the dust
in the local universe. We conclude that the polarimetry supports the view
that the average dust-to-gas ratio in the GRB host galaxy along the line 
of sight is much lower than those in our Galaxy.

\section{Discussion}

\subsection{Limits on the host and a supernova component \label{host}}

On our VLT images and in the afterglow light curve we do not find evidence for
an underlying host galaxy (Fig.~\ref{hosts}). Spectroscopic evidence for
emission lines from an underlying host is missing too (although mainly due to
the insufficient wavelength coverage of our spectra). Based on our latest
$R$-band imaging on March 13, i.e., 15\,days after the burst, we conclude that
$R_{\rm host}>26.2$. This estimate does not exclude the possibility that the
host is indeed brighter, but some arcsec away from the burster (e.g., the
objects 1 and 2 in Fig.~\ref{hosts}), for which we have no spectroscopic
confirmation.

At the given redshift of $z$=1.986, we do not expect to see an underlying
supernova (SN) component in our data: Assuming SN 1998bw as a template (for
details, see Zeh et al. 2004), this predicts an unabsorbed peak magnitude of
$R\approx28$ approximately at day 33 and a negligible contribution to the
afterglow light within the first week after the burst. That neither an
underlying host galaxy nor a SN component contributes substantially to the
light of the optical transient, even 1 week after the burst, means that we
essentially observed an uncontaminated GRB afterglow.

\subsection{The nature of the afterglow \label{afterglow}}

An achromatic break in the light curve is generally considered as evidence for
a jetted explosion (Rhoads 1999; M\'esz\'aros \& Rees 1999). Using standard
procedures, for the given break time and redshift as well as
gamma-ray fluence the beaming corrected energy output of GRB 030226 is
consistent with the ensemble statistic of bursts; the results we obtain based
on our deduced break time is basically the same as in Pandey et al. (2004),
and we refer the reader to this paper for a more detailed discussion on this
subject.

A non-spherical explosion in the form of a highly relativistic jet should
produce net polarization, and this is indeed what has been observed in a
number of afterglows so far (for a review, see Covino et
al. 2004). Geometrically simple models (e.g., Ghisellini \& Lazzati 1999; Sari
1999) assume a uniform jet, with a constant energy per unit solid angle
accross the full jet. In this case, the polarization evolution has two peaks,
separated by an epoch of null polarization, which roughly coincides with the
break in the light curve. The polarization angle is expected to flip by
90$\degr$ between the two peaks. More complex models (e.g., Rossi et al. 2004;
Lloyd-Ronning et al. 2004) assume instead an energy distribution per solid
angle which decreases towards the edge of the jet. In this case, even if the
light curves are not very different from the uniform case, the polarization is
expected to show a peak close to the time of the break. Our polarization
measurement  of the afterglow of GRB 030226 was indeed performed very close to
the break in the light curve. Interpreting this break as due to the jet
effect, our low polarization value would thus favor a uniform jet  over a
structured jet. However, no solid conclusion can be drawn from a single upper
limit, lacking monitoring of the polarization level and angle evolution with
time.

\begin{figure}
\plotone{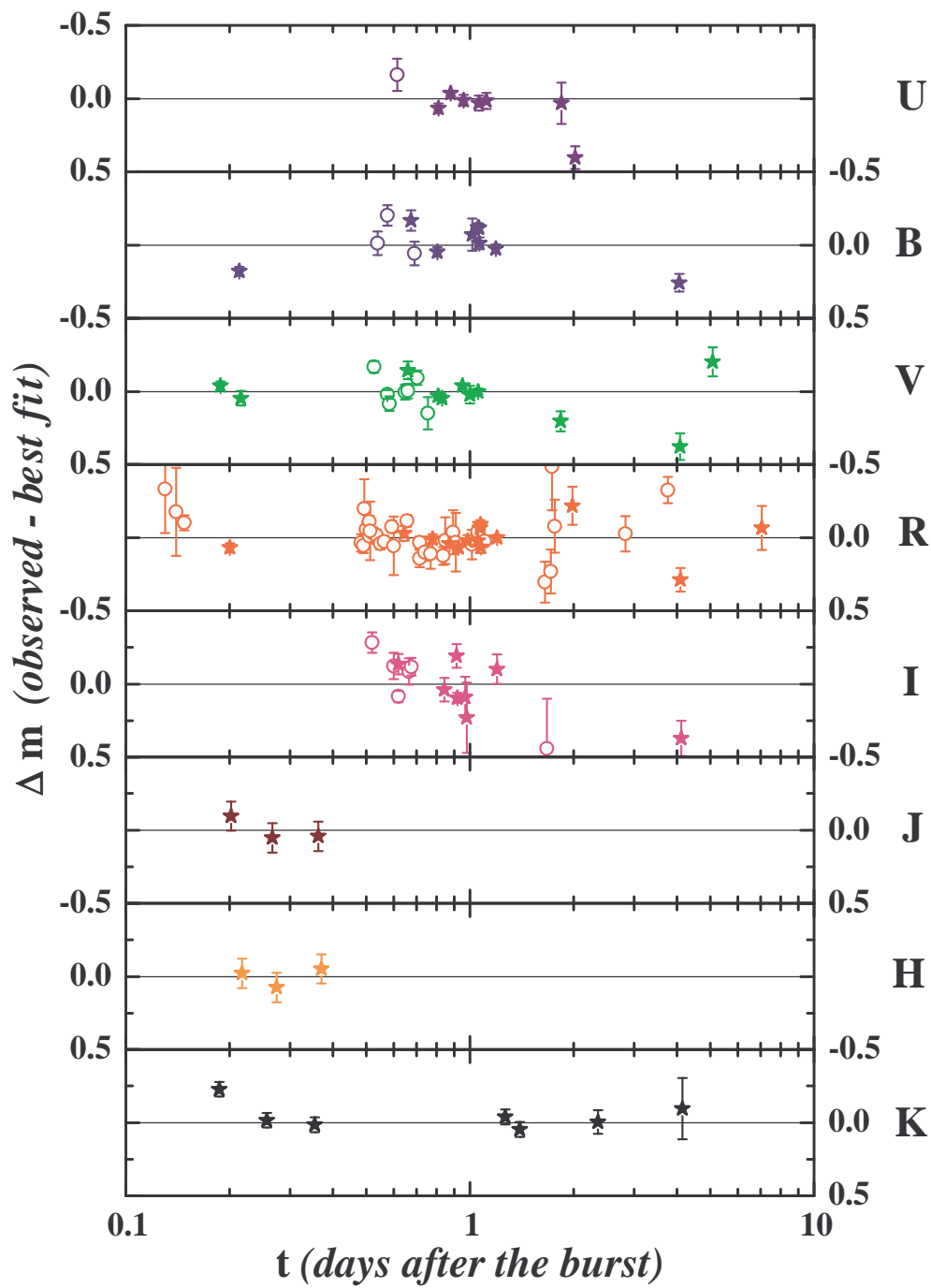}
\caption{Observed minus best fit magnitudes
(based on the $R$-band fit;  Fig.~\ref{Rband}) for the data from
Table~\ref{log} together with data reported in several GCNs 
(for references, see Fig.~\ref{Rband}) and by Pandey et
al. (2004). The scale at the left ordinate refers to $U,V,I, H$, the right to
$B,R,J, K$. $\Delta m < 0$ means an excess of flux with respect to the fit.
Symbols follow Fig.~\ref{Rband}.}
\label{OC}
\end{figure}

\begin{figure}
\includegraphics[scale=.60,angle=-90]{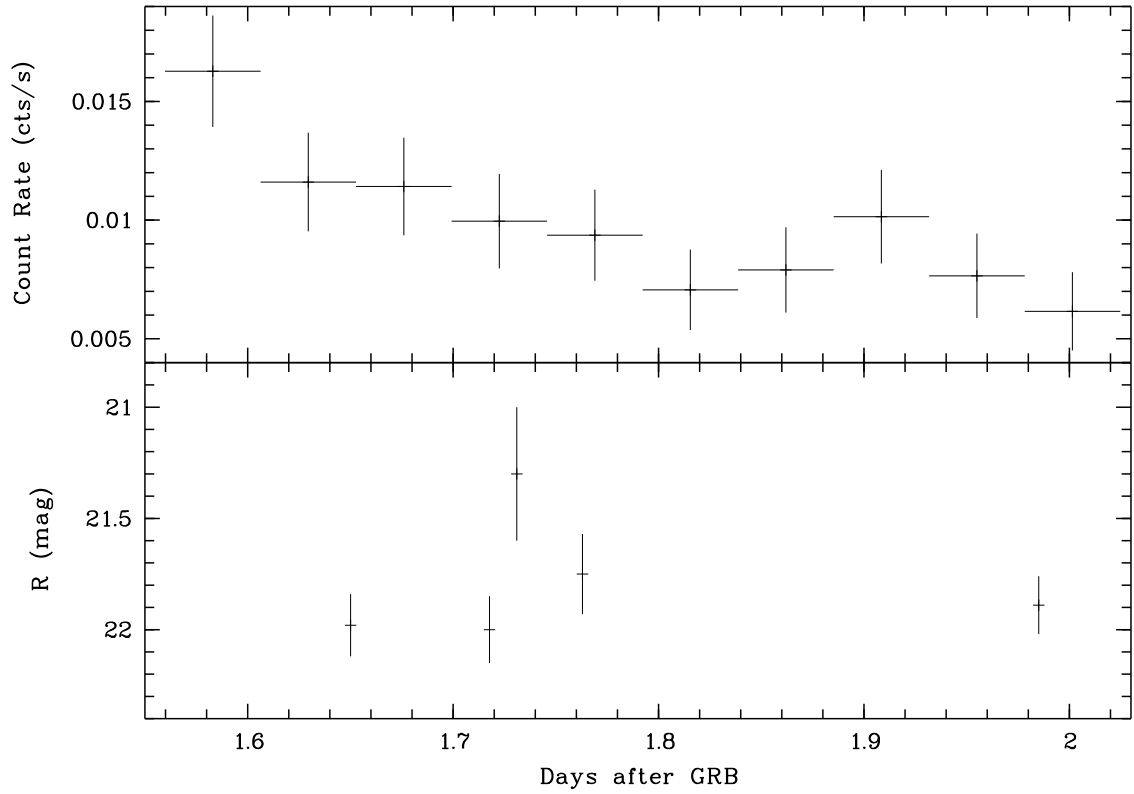} 
\caption{X-ray light curve obtained with {\it Chandra} (top panel) and 
the contemporaneous optical observations (bottom panel) from
Fatkhullin et al. (2003), Semkov (2003), Pandey et al. (2004), and
our observing run (Table~\ref{log}).}
\label{xlc} 
\end{figure}

\begin{figure}
\includegraphics[scale=.60,angle=-90]{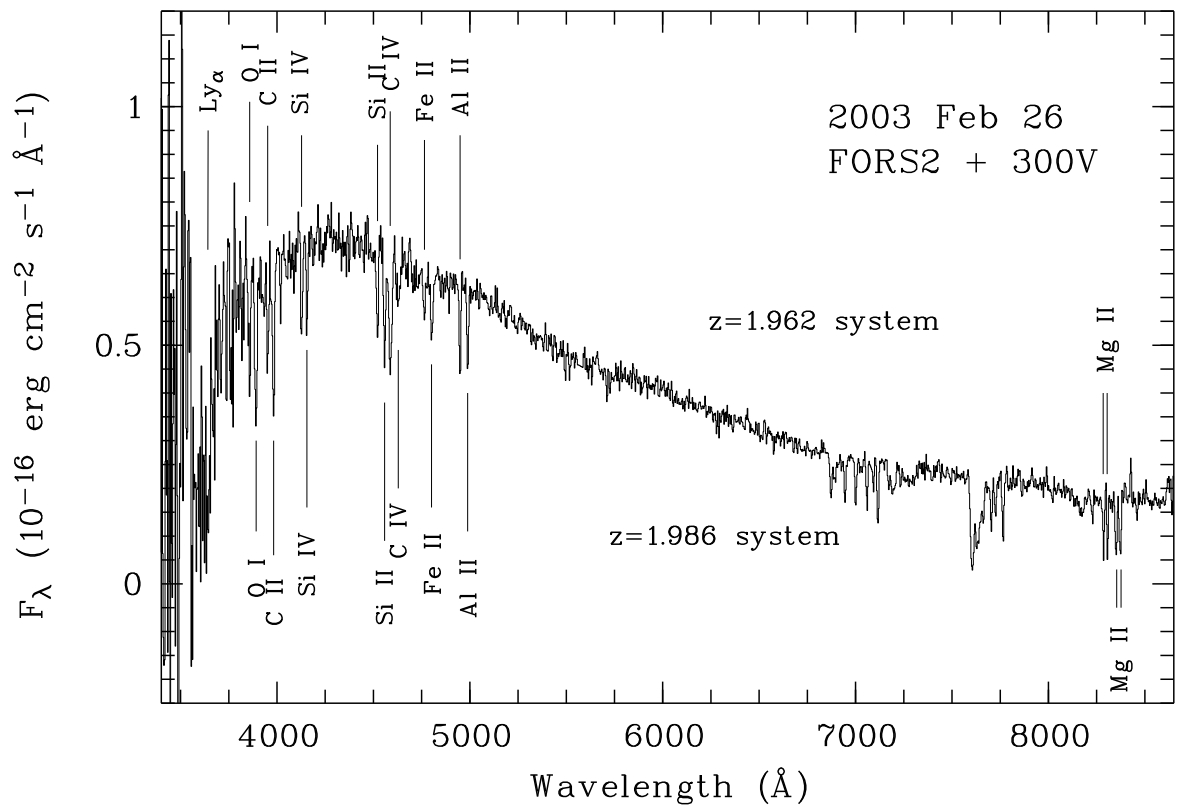} 
\caption{Short-wavelength part of the VLT/FORS2 spectrum using the 300V grism 
taken on 2003 February 26, five hours after the burst. Two distinct redshift
systems at $z$=1.962 and $z$=1.986 are detected. Line identifications are
indicated. Note the broad Ly~${\alpha}$ absorption feature. Some telluric
lines between $\sim$6500 and  8000~\AA \ could not be removed during the
calibration process. Also, the rather unexpected curvature in the blue
wavelength range is probably unreal and may be a residual of the
flux calibration.}
\label{fig:forsNight1} 
\end{figure}

\begin{figure}
\includegraphics[scale=.60,angle=-90]{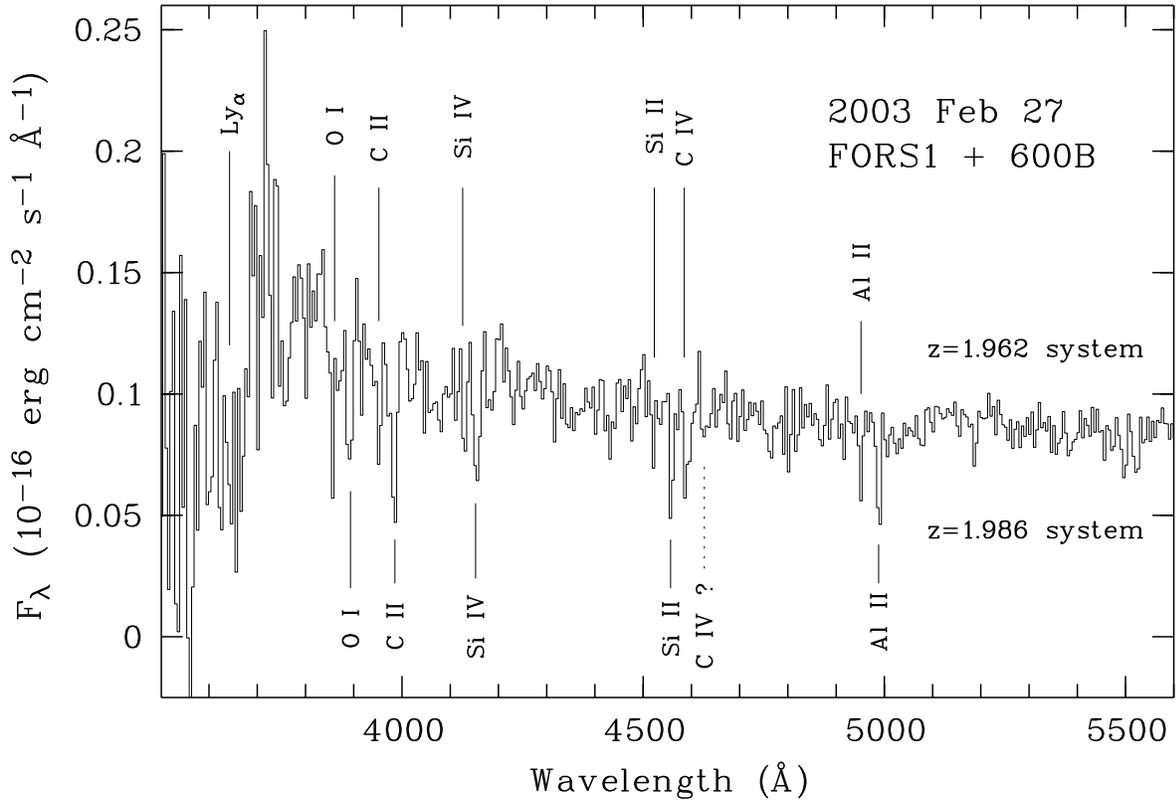} 
\caption{VLT/FORS1 spectrum from the GRB afterglow using the 600B grism
taken on 2003 February 27, twenty nine hours after the burst and thus
shortly after the break in the light curve.}
\label{fig:forsNight2} 
\end{figure}

\begin{figure}
\includegraphics[scale=0.9,angle=0]{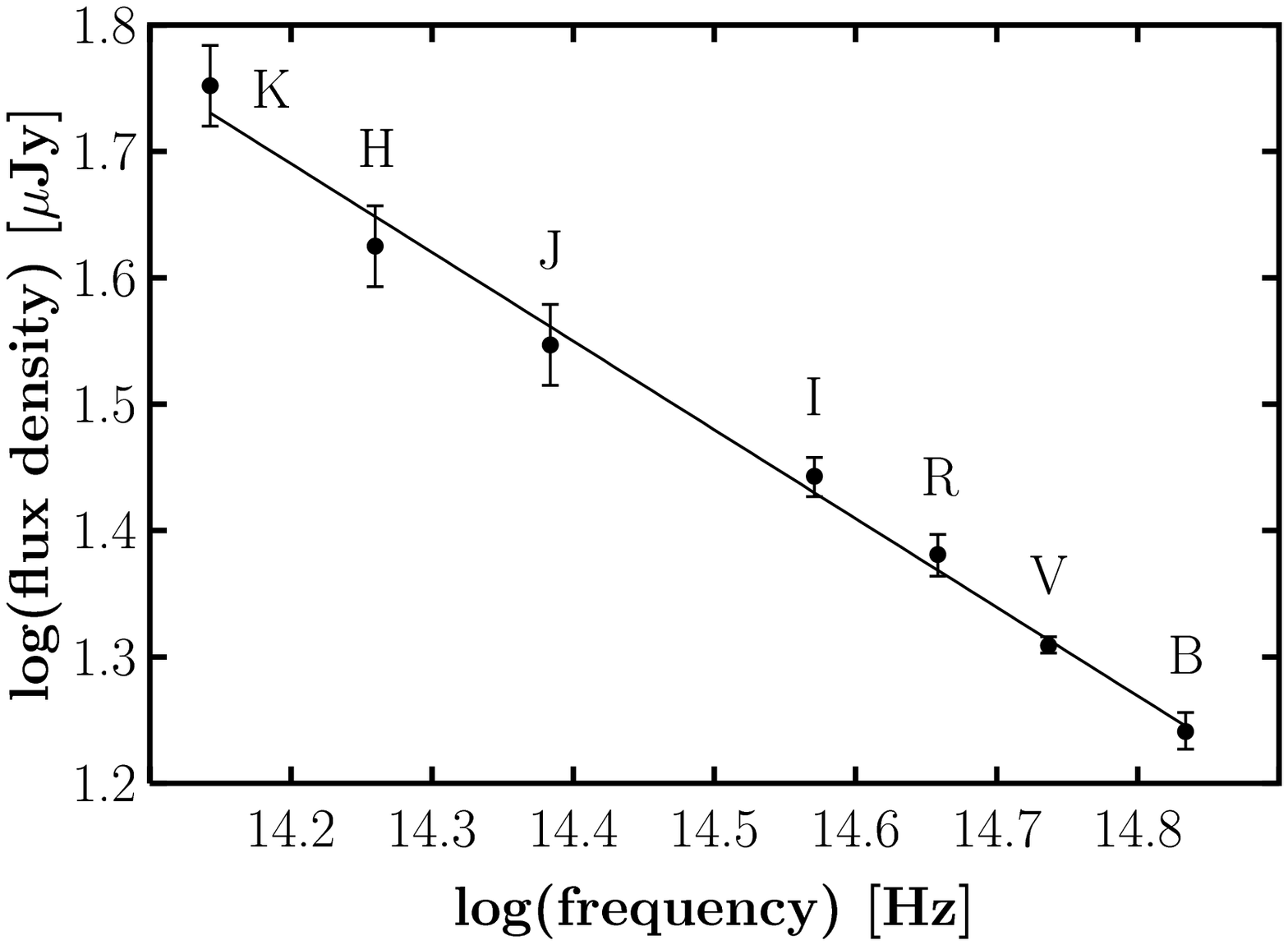}
\caption{Spectral energy distribution of the afterglow at the time of the 
light curve break ($t_b$; \S~\ref{alphas}).  The data are corrected for 
Galactic extinction.}
\label{SED}
\end{figure}

\begin{figure}
\includegraphics[scale=.60,angle=-90]{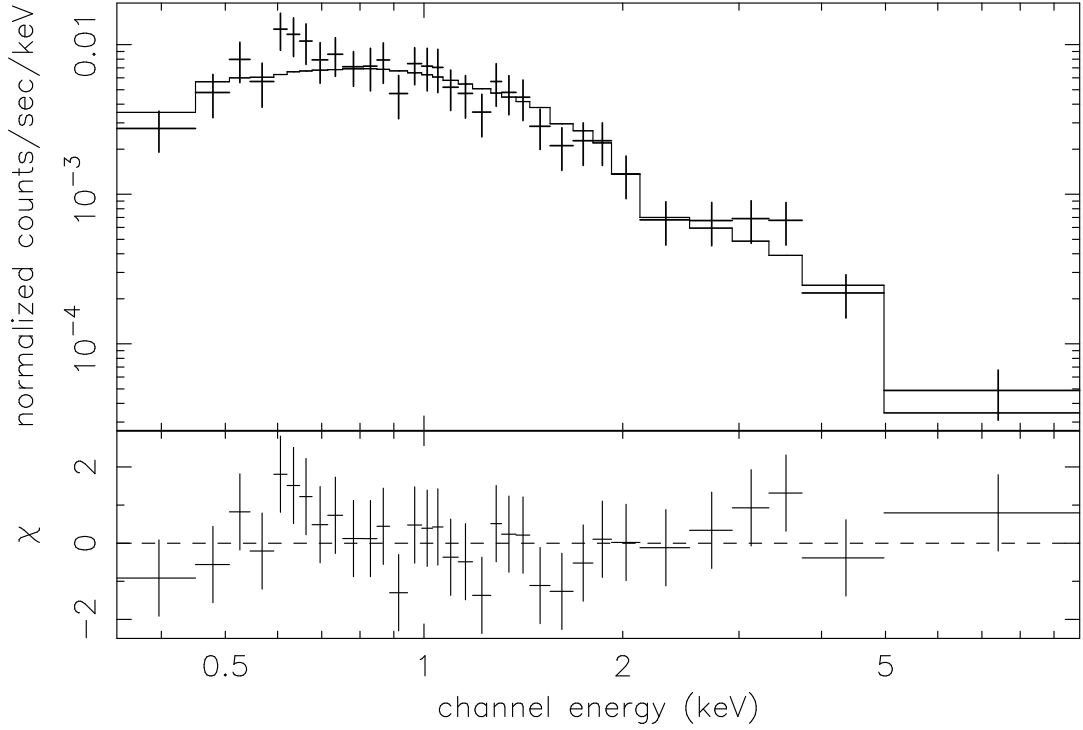}
\caption{The X-ray count spectrum (top) of the GRB 030226 afterglow, integrated
over the full observation time (1.54--2.04 days after the GRB).
A power-law model (full line) gives a reasonably accurate fit 
($\chi^2$/d.o.f.=0.7; deviations from the power law fit in units
of $\sigma$ are shown in the bottom panel) with a spectral slope 
$\beta_X=1.04\pm0.20$.} 
\label{xsp}  
\end{figure}

\begin{figure}
\plotone{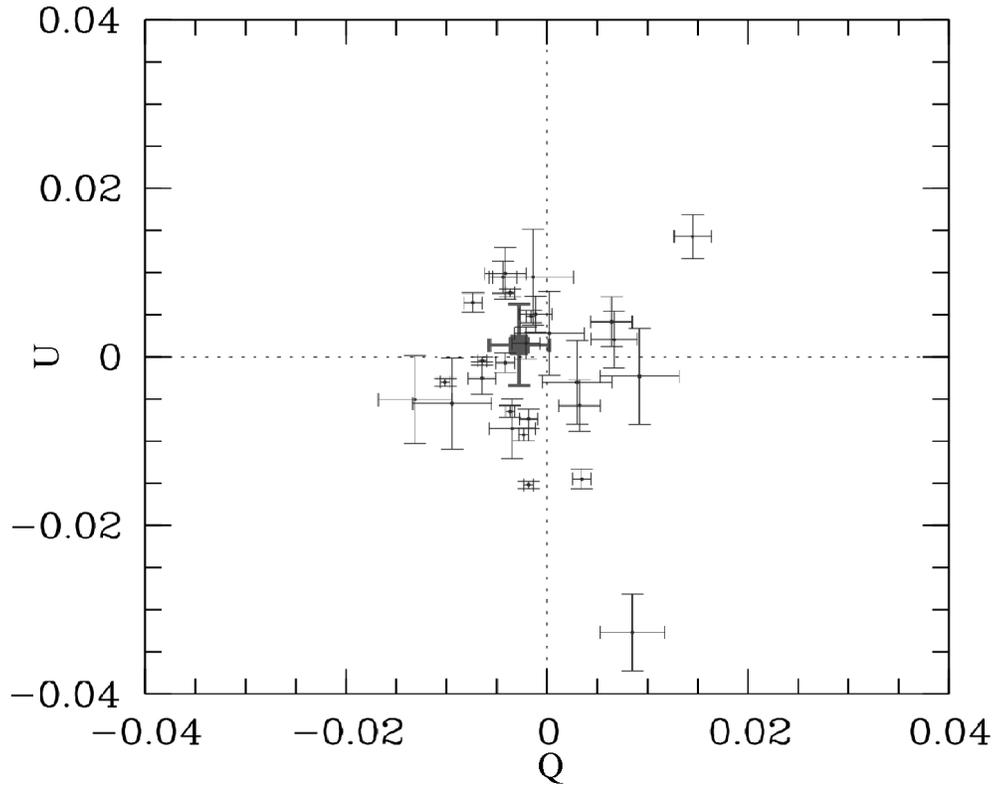}
\caption{Stokes $Q$-$U$-plane showing the position of the afterglow (big dot) 
compared with that of field stars. The average polarization of field stars is 
also close to zero, as it is expected towards such a low-reddening line of 
sight.}
\label{QU}
\end{figure}

\begin{figure}
\plotone{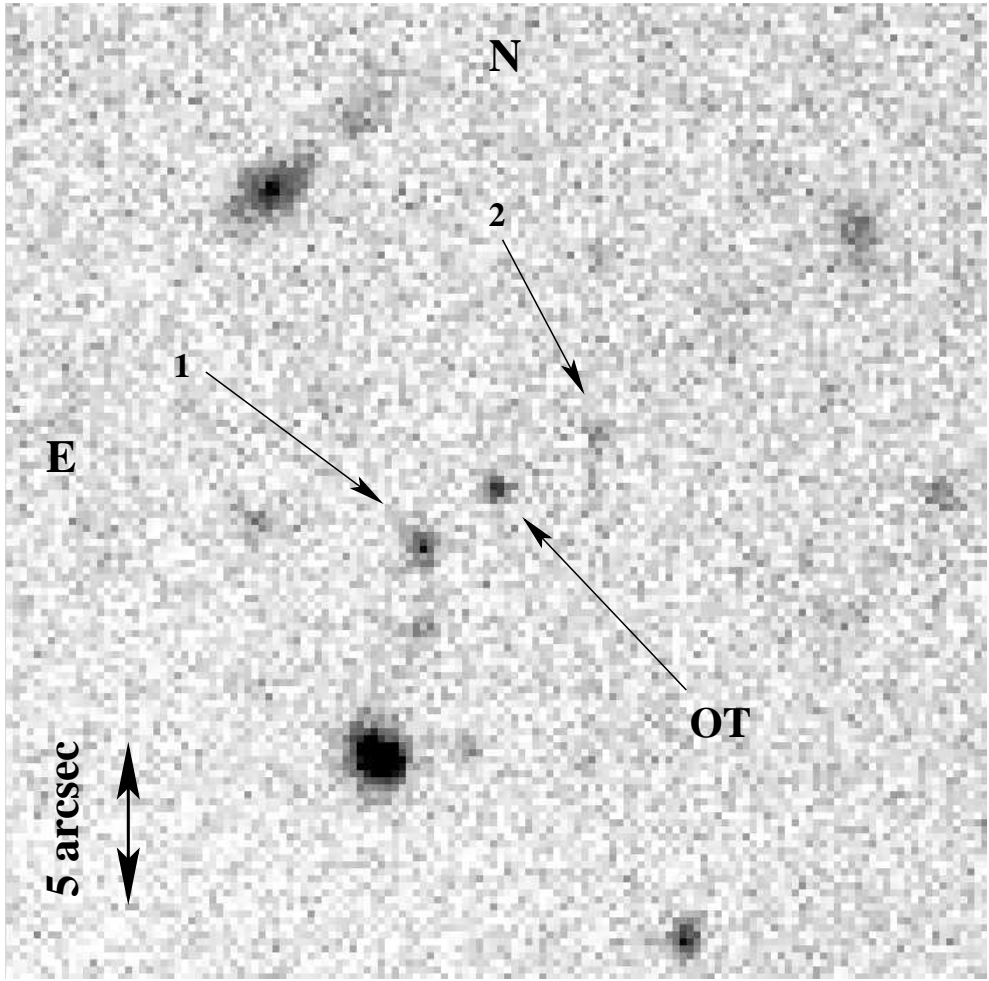}
\caption{VLT/FORS1 $R$-band image taken on March 2, four days after the 
burst. At that time the magnitude of the optical afterglow was $R$=24.2. There
are two sources within 2$\farcs$5 radius around the optical transient
(indicated by numbers and arrows). Without having spectral data at hand, we
cannot exclude that one of them is the host. However, in such a case the GRB
would have occurred at a rather large projected distance from the luminosity
center of the host: At $z$=1.986 an angular distance of 1$''$  corresponds to
9 kpc. It is conceivable that one of these sources is responsible  for the
foreground absorber of Mg II at $z$=1.042 we found in our spectra
(\S~\ref{cosmology}).}
\label{hosts}
\end{figure}

An inspection of Fig.~\ref{OC} shows that all photometric data concentrate
around the break time, while the other part of the light curve is more poorly
sampled. Depending on the extent to which we merged data taken from different
groups at different optical telescopes, our results vary between strong and
weak evidence for a short-term variability in the light curve. These
potential short-term fluctuations, primarily at later times, are reflected in
the relatively large 1$\sigma$ error bars of the deduced light curve
parameters $\alpha_1$ and $\alpha_2$. In particular, on day 4 after the burst,
the afterglow flux is clearly below the expectations based on the light curve
fit, and this holds for all optical bands (Fig.~\ref{OC}). Unfortunately, this
is the only epoch after the break time where we have a homogeneous multi-color
data set. Alternatively, there could be an excess of flux at later times
($t>5$ days), which cannot be attributed to an underlying host or supernova,
however. In this context it is not surprising that Pandey et al. (2004)
derived a different post-break decay slope of $\alpha_2 =
2.05\pm0.04$ given the potential flux variations of the afterglow in
combination with their much smaller monitoring time.

A key parameter of the afterglow SED is the location of the cooling
frequency $\nu_c$, which separates fast-cooling from slow-cooling electrons
(Sari et al. 1998). The combination of the \it Chandra \rm X-ray data with the
optical/NIR data places $\nu_c$ between the X-ray and the optical bands, in
agreement with the conclusion drawn by Pandey et al. (2004). Thereby the
spectral slopes are similar to those found for other afterglows (e.g., GRB
010222: Masetti et al. 2001; GRB 011211: Jakobsson et al. 2003). Standard
fireball models for jetted explosions with $p>2$ (Sari et al. 1998, 1999;
Livio \& Waxman 2000; Chevalier \& Li 2000; Dai \& Cheng 2001) then predict
$p=2\beta+1$ and $\alpha_2=p$ (i.e., $\alpha_2 = 2.40\pm0.06$ for $\beta =
0.70\pm0.03$; \S~\ref{SEDsect}), consistent with the observed late-time decay
slope of the afterglow. 

\begin{figure}
\includegraphics[scale=0.85,angle=0]{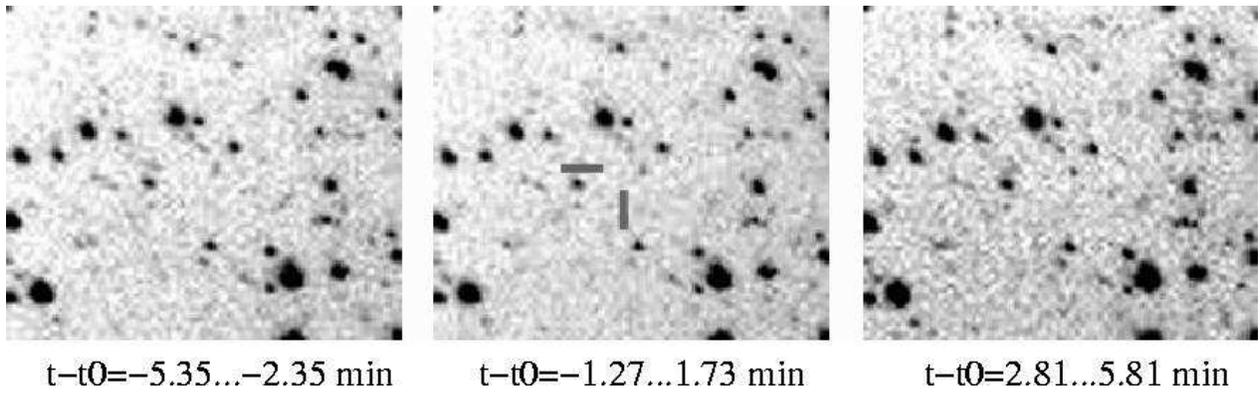}
\caption{Unfiltered exposures (180-s each) obtained by the
BOOTES-1 wide field camera covering the GRB 030226 error box. The images
covered the period 03:41 - 03:52 UT on 26 Feb 2003 
(i.e., between 5.3 min before
the trigger and 5.8 min after the trigger). The ticks mark the position of the
optical afterglow. The despicted field is 2$\times$2 deg$^2$. Note that there 
is some degree of geometrical distortion due to the fact that the images were 
taken with a wide-field optical system. North is up and east to the left.}
\label{bootes}
\end{figure}

Based on the above discussion we conclude that the afterglow of GRB 030226
showed all the properties already known from other bursts: an achromatic
break around 1 day after the burst very likely due to a jetted outflow, a
rapid decay thereafter, and possible short-term fluctuations. Finally, we
can use the BOOTES-1 observations to constrain the optical emission before,
during, and shortly after the genuine GRB event. A visual inspection of all
BOOTES-1 frames reveals no such emission at the position of the optical
afterglow, in particular simultaneously to the burst itself
(Fig.~\ref{bootes}). Therefore we derive an upper limit of $R$ = 11.5 (due to
cirrus clouds present in the sky at the time of the event) for the reverse
shock emission (if any) arising from this GRB, in contrast to the 9-th
magnitude prompt flash emission recorded for GRB 990123 (Akerloff et al. 1999)
which demonstrates that a bright, prompt optical emission is not a generic
characteristic of GRBs.

\subsection{Clues on the GRB progenitor}

Of special interest are the highly ionized absorbers Si\,IV and C\,IV  and the
associated absorption line systems seen in our VLT spectra separated in
velocity by 2400 km s$^{-1}$. Schaefer et al. (2003) and Mirabal et al. (2003)
discussed a similar finding for the afterglow of GRB 021004 ($z$=2.3). The
afterglow of this burst not only showed these lines, but four absorption line
systems separated by 450, 990 and 3155 km s$^{-1}$ (Mirabal et al. 2003;
Castro-Tirado et al. 2004). Similar features were found in the afterglow of
GRB 020813 (Barth et al. 2003). Schaefer et al. (2003) and Mirabal et
al. (2003) have provided strong arguments that these lines, together with the
observed absorption line systems, are likely to come from expanding shells
around a massive Wolf-Rayet star (cf. Nugis \& Lamers 2000; for a discussion,
see also Wijers 2001; Chevalier et al. 2004). This could also explain the
strong flux variations seen in some afterglows (e.g., GRB 000301C: Garnavich
et al. 2000; Masetti et al. 2000; GRB 011211: Jakobsson et al. 2003; GRB
021004: Bersier et al. 2003; Holland et al. 2003; Lazzati et al. 2002; GRB
030329: Matheson et al. 2003), which might also be apparent for GRB 030226,
even though with lower amplitude (Fig.~\ref{OC}). This is consistent with
theoretical models that link long-duration GRBs to the explosions of massive
stars (e.g., Fryer et al. 1999; Paczy\'nski 1998; Heger et al. 2003).

For GRB 030226 the $z$=1.986 system is clearly not a scaled version of the
$z$=1.962 system. Basically, we identify two groups of ions. Those of Si\,IV,
Fe II, and Al II have roughly the same rest-frame equivalent widths in both
redshift systems, whereas those of the other species are approximately twice
as abundant in the $z$=1.986 system compared to the $z$=1.962 system. Within
the context of the Wolf-Rayet star model this would reflect different chemical
compositions of different shells in the wind, a phenomenon that is observed
for example in the outer ejecta of Eta Carina (Smith \& Morse 2004). While
evidence for 'line locking' has been found in spectra taken from the afterglow
of GRB 021004 (Savaglio et al. 2002; M{\o}ller et al. 2002; see also the
discussion in Mirabal et al. 2003), this phenomenon is not apparent in our
spectra of the afterglow of GRB 030226. If not hidden because of the medium
spectral resolution, this implies that the observed features in both
absorption line systems are not due to radiative acceleration of the absorbing
clouds.

In principle, the presence of a strong Ly\,$\alpha$ absorber in the spectrum
of the afterglow of GRB 030226 implies a dense interstellar medium, possibly
at the redshift of the burster. However, in our spectra the  redshift of the
Ly\,$\alpha$ line cannot be accurately determined. We can only place the
constraint 1.93$<z<$1.99. Similarly, a potential variability of the
absorption lines is of interest since it could place the absorbing clouds
close to the burster (e.g., Perna \& Loeb 1998; Draine 2000). However, a
comparison of our medium-resolution VLT spectra taken 0.2 and 1.2 days after
the burst does not reveal evidence for a general variation in the line
strengths. Unfortunately, the Ly\,$\alpha$ line is difficult to analyze in
this respect given its unfavorable position at the very blue side in our VLT
spectra. 

While our spectral data support the view that the GRB progenitor was a massive
star, there is no evidence for a stellar wind profile in the light curve data,
similar to other cases (cf. Chevalier \& Li 2000; Panaitescu \& Kumar 2002).
Within the framework of standard afterglow models with $p>2$ the light curve
parameter $\alpha_1$ is related to the density profile of the circumburst
medium, $n(r)\sim r^{-k}$, by $\alpha_1 = 3/4\,(p-1)+k/(8-2k)$ (e.g.,
M\'esz\'aros, Rees, \& Wijers 1998; Dai \& Wu 2003). In the ideal case $k=0$
means an ISM-shaped interstellar medium, while $k=2$ refers to a stellar
wind. Using $\alpha_1=0.50\pm0.35$ and $p>2$ a density profile with $k$=2 is
basically ruled out. Also the rather steep change in the decay slope between
0.5 and 2 days after the burst from $\alpha_1=0.50\pm0.35$ to $\alpha_2 =
2.55\pm 0.38$ might be difficult to explain with a standard wind profile. But
does this really argue against WR stars as GRB progenitors? There is
increasing observational evidence that the outer winds of these stars might
indeed be very clumpy (e.g., Crowther et al. 2002) with the wind properties
depending on the viewing direction, comparable to the one of Eta Carina (see
Fig.~1 in van Boekel et al. 2003; Smith et al. 2003). An ansatz with $k=2$ to
describe the stellar wind might then be an oversimplification; in reality $k$
might be notably smaller (Wijers 2001). More insight on the properties  of
winds from WR stars and more spectroscopic  afterglow data are required in
order to tackle this problem. A monitoring of the evolution of the cooling
frequency across the optical/NIR bands in GRB afterglows, which might become
possible in the \it Swift \rm era (Gehrels 2004),  might also help to clarify
this issue.

\section{Summary}

We have presented a comprehensive data set of the afterglow of GRB 030226,
consisting of optical, NIR, and X-ray observations, including VLT
spectroscopy and VLT polarimetry. Concerning the nature of the GRB afterglow
our basic findings are the following:
\begin{enumerate}
\item The evolution of the afterglow was achromatic  within the observational
errors over the optical/NIR bands analogous to other bursts (e.g., GRB
000301C: Jensen et al. 2001;  GRB 020405: Masetti et al. 2003). A break in the
light curve occurred $\sim$ 1 day after the burst. The break was rather smooth,
similar to that found for, e.g., GRB 990510 (Harrison et al. 1999).
\item Because the break was achromatic we interpret it as due to a jetted
explosion. Close to the break time the intrinsic linear polarization of the
afterglow was very low ($<1.1\%$), which favors a uniform jet over a
structured jet. 
\item After correction for Galactic extinction, the best fit multi-color
lightcurve has a spectral slope across the optical/NIR bands of
$\beta=0.70\pm0.03$, with no sign of additional reddening due to dust in the
host galaxy. This is in contrast to the \it Chandra \rm X-ray data, which
point to a relatively large hydrogen column density along the line of sight. 
This perhaps indicates a reduced dust-to-gas ratio in the GRB environment, 
similar to suggestions made for several other GRB afterglows (Galama
\& Wijers 2001; Hjorth et al. 2003b; Stratta et al. 2004).
\item In the 0.3--10 keV region the spectral slope was $\beta=1.04\pm0.20$ on
day 1.6--2, which together with the spectral slope across the optical/NIR
bands places the cooling frequency between the optical bands and the X-ray
region.
\end{enumerate}
In addition, based on BOOTES-1 observations no prompt optical flash coincident
with the gamma-ray burst was  detected. We place an upper limit of $R$ = 11.5
on the reverse shock emission  arising from this burst.

While our multi-wavelength observations trace the evolution of the afterglow,
our VLT spectroscopy provides insight on the nature of the GRB environment,
even though an underlying host galaxy remained undetected down to a faint flux
level ($R>26.2$). In particular, beside a foreground absorber at $z$=1.042 we
found two absorption line systems at redshifts $z$=1.962 and $z$=1.986, with
the higher value as a lower limit for the redshift of the burster. Both line
systems reveal an interstellar gas composed of highly ionized species (Si\,IV,
C\,IV). This, together with the inferred separation in velocity of 2400 km
s$^{-1}$ between both systems, is very similar to observations of the
afterglow of GRB 021004 (Castro-Tirado et al. 2004; Schaefer et al. 2003;
Mirabal et al. 2003), and possibly indicative of a fast stellar wind placed in
a strong radiation field. Within this context the data support an association
of GRBs with Wolf-Rayet stars. They also call for early spectroscopy of as
many GRB afterglows as possible in order to search for potential line features
from stellar winds. It is surely premature to claim that all GRB progenitors
are associated with WR stars, but among all progenitor models discussed so far
in the literature (cf. Fryer et al. 1999) they are most promising from a
theoretical point of view, and significantly supported by present-day
observations.

\section{Acknowledgments}

We are indebted to the ESO staff at La Silla and at Paranal, in particular
Malvina Billeres, Lisa Germany, Swetlana Hubrig, Nuria Hu\'elamo, Norma
Hurtado, Elena Mason, Leonardo Vanzi for prompt execution  of the observing
requests and additional efforts related to that.  We also thank H. Navasardyan
for the coordination and execution of the  Asiago observations in Service
Mode. J.G. and K.P. are particularly grateful to H. Tananbaum for granting the
\it Chandra \rm observing time.  S.K. acknowledges support from the Deutsche
Akademische Austauschdienst (DAAD) under grants No. D/0237747 and D/0103745
and from the Deutsche Forschungsgemeinschaft (DFG) under grant Kl766/11.
N.M., both E.P., and E.M. acknowledge support under CRUI Vigoni program
31-2002. D.H.H. acknowledges support under NSF grant
INT-0128882. A.Z. acknowledges financial support from the
Friedrich-Schiller-University, Jena, Germany, and from the DFG under grant
Kl766/11. K.P. acknowledges support from the Carlsberg foundation. A.J.C.-T.
acknowledges support from INTA and  the Spanish Programme AYA 2004-01515 as
well as the encouraging work of T. J. Mateo Sanguino, P. Kub\'anek and
S. Vitek. This research has made use of the NASA/IPAC Infrared Science
Archive, which is operated by the Jet Propulsion Laboratory, California
Institute of Technology, under contract with the National Aeronautics and
Space Administration. We thank the anonymous referee for valuable comments.

\newpage

\appendix

\section{AB offsets}

\begin{table}[h] 
\centering
\caption[]{Conversion constants for the transformation\\
into the AB magnitude system (Oke \& Gunn 1983).\tablenotemark{a}}
\label{ABmags}
\begin{tabular}{llrr} 
\noalign{\smallskip} \hline \noalign{\smallskip} 
Instrument  & Filter   & eff. wavelength & AB$_{\rm off}$  \\
\noalign{\smallskip} \hline \noalign{\smallskip} 
& & & \\
VLT/FORS1   & $B_-$Bess +34 & 429 nm   & --0.10 \\
VLT/FORS1   & $V_-$Bess +35 & 554 nm   &   0.03 \\
VLT/FORS1   & $R_-$Bess +36 & 657 nm   &   0.23 \\
VLT/FORS1   & $I_-$Bess +37 & 768 nm   &   0.44 \\
            &               &          &        \\[-2mm]
VLT/FORS2   & $V_-$Bess +75 & 554 nm   &   0.06 \\
            &               &          &        \\[-2mm]
VLT/ISAAC   & $J_s$         & 1240  nm &   0.90 \\
VLT/ISAAC   & $H$           & 1650  nm &   1.38 \\
VLT/ISAAC   & $K_s$         & 2160  nm &   1.86 \\
            &               &          &        \\[-2mm]
NTT/EMMI    & $B$ \#605     & 413.9 nm & --0.16 \\
NTT/EMMI    & $V$ \#606     & 542.6 nm &   0.02 \\
NTT/EMMI    & $R$ \#608     & 641.0 nm &   0.22 \\
            &               &          &        \\[-2mm]
NTT/SuSI2   & Bessel $B$ \#811 & 421.2 nm & --0.06 \\
NTT/SuSI2   & Bessel $R$ \#813 & 641.6 nm &   0.21 \\
            &               &          &        \\[-2mm]
NOT/ALFOSC  & $U$ \#7       & 362 nm   &   0.72 \\
NOT/ALFOSC  & $V$ \#75      & 530 nm   &   0.01 \\
            &               &          &        \\[-2mm]
TNG/Dolores & $U$ \#01      & 361 nm   &   0.78 \\
\noalign{\smallskip} \hline
\end{tabular}
\tablenotetext{a}{
For the VLT/ISAAC filters the AB offset [mag] was taken from Labb\'e et al. 
(2003) and Gorosabel et al. (2003). For the other cases AB offsets were 
calculated in the present work using the effective wavelengths 
and detector quantum efficiencies as provided at the webpages of the 
corresponding observing sites.}
\end{table}


\clearpage

\clearpage

\clearpage

\clearpage

\clearpage

\clearpage

\clearpage

\clearpage

\clearpage

\clearpage

\clearpage

\clearpage

\clearpage

\begin{deluxetable}{lcllcr}
\tablecolumns{6}
\tablewidth{0pc}
\tablecaption{Log of the observations\label{log}} 
\tablehead{
\colhead{~~~~~~~Date (UT)} &
\colhead{$<dt>$ (days)\tablenotemark{a}} &
\colhead{Tel./Instr.} &
\colhead{Filter/Grism} &
\colhead{Exp. (s)} &
\colhead{magnitude}}
\startdata
 2003 Feb 26, 08:04 & 0.1868 & VLT/ISAAC  & $K_s$ & 15$\times$10 & 15.95$\pm$0.05\\
 2003 Feb 26, 08:09 & 0.1882 & NTT/EMMI   & $V  $ &  3$\times$300& 19.09$\pm$0.03\\
 2003 Feb 26, 08:27 & 0.2007 & NTT/EMMI   & $R  $ &  3$\times$300& 18.87$\pm$0.03\\
 2003 Feb 26, 08:29 & 0.2024 & VLT/ISAAC  & $J_s$ & 10$\times$30 & 17.56$\pm$0.10\\
 2003 Feb 26, 08:43 & 0.2062 & VLT/FORS2  & $V$   &  1$\times$10 & 19.27$\pm$0.05\\
 2003 Feb 26, 08:46 & 0.2139 & NTT/EMMI   & $B  $ &  3$\times$300& 19.76$\pm$0.03\\
 2003 Feb 26, 08:52 & 0.2173 & VLT/FORS2  & 300V  &  1$\times$900& \nodata \\
 2003 Feb 26, 08:49 & 0.2180 & VLT/ISAAC  & $H  $ & 15$\times$12 & 17.02$\pm$0.10\\
 2003 Feb 26, 09:51 & 0.2569 & UKIRT/UFTI & $K  $ &  9$\times$60 & 16.39$\pm$0.05\\
 2003 Feb 26, 10:04 & 0.2666 & UKIRT/UFTI & $J  $ & 11$\times$60 & 17.91$\pm$0.10\\
 2003 Feb 26, 10:17 & 0.2746 & UKIRT/UFTI & $H  $ &  9$\times$60 & 17.29$\pm$0.10\\
 2003 Feb 26, 12:12 & 0.3545 & UKIRT/UFTI & $K  $ &  9$\times$60 & 16.69$\pm$0.05\\
 2003 Feb 26, 12:23 & 0.3621 & UKIRT/UFTI & $J  $ &  9$\times$60 & 18.16$\pm$0.10\\
 2003 Feb 26, 12:34 & 0.3698 & UKIRT/UFTI & $H  $ &  9$\times$60 & 17.42$\pm$0.10\\
 2003 Feb 26, 18:19 & 0.6194 & Wendelstein& $I  $ &  3$\times$600& 19.37$\pm$0.07\\
 2003 Feb 26, 19:04 & 0.6444 & Wendelstein& $R  $ &  2$\times$600& 19.90$\pm$0.05\\
 2003 Feb 26, 19:27 & 0.6604 & Wendelstein& $V  $ &  2$\times$600& 20.19$\pm$0.06\\
 2003 Feb 26, 19:50 & 0.6746 & Wendelstein& $B  $ &  2$\times$600& 20.56$\pm$0.07\\
 2003 Feb 26, 22:29 & 0.7795 & JKT        & $R  $ & 1$\times$1200& 20.22$\pm$0.03\\
 2003 Feb 26, 22:43 & 0.8035 & Asiago     & $B  $ & 2$\times$1200& 21.04$\pm$0.03\\
 2003 Feb 26, 23:07 & 0.8076 & NOT/ALFOSC & $V  $ &  1$\times$300& 20.67$\pm$0.02\\
 2003 Feb 26, 23:13 & 0.8104 & NOT/ALFOSC & $U  $ &  1$\times$600& 20.81$\pm$0.03\\
 2003 Feb 26, 23:26 & 0.8298 & Asiago     & $V  $ &  2$\times$900& 20.73$\pm$0.02\\
 2003 Feb 26, 23:38 & 0.8437 & Tautenburg & $I  $ & 15$\times$120& 20.01$\pm$0.08\\
 2003 Feb 26, 23:59 & 0.8726 & Asiago     & $R  $ &  2$\times$600& 20.44$\pm$0.03\\
 2003 Feb 27, 00:43 & 0.8795 & TNG        & $U  $ & 1$\times$1200& 20.84$\pm$0.02\\
 2003 Feb 27, 01:21 & 0.9153 & Tautenburg & $I  $ & 15$\times$120& 19.92$\pm$0.08\\
 2003 Feb 27, 01:40 & 0.9201 & Asiago     & $I  $ &  2$\times$600& 20.21$\pm$0.03\\
 2003 Feb 27, 02:02 & 0.9274 & JKT        & $R  $ & 1$\times$1200& 20.57$\pm$0.03\\
 2003 Feb 27, 02:35 & 0.9521 & NOT/ALFOSC & $V  $ &  1$\times$300& 20.88$\pm$0.02\\
 2003 Feb 27, 02:42 & 0.9587 & NOT/ALFOSC & $U  $ &  1$\times$600& 21.04$\pm$0.04\\
 2003 Feb 27, 02:49 & 0.9677 & Wendelstein& $I  $ &  2$\times$600& 20.30$\pm$0.14\\
 2003 Feb 27, 03:08 & 0.9785 & Tautenburg & $I  $ &  5$\times$120& 20.46$\pm$0.24\\
 2003 Feb 27, 03:12 & 0.9837 & Wendelstein& $R  $ &  2$\times$600& 20.63$\pm$0.04\\
 2003 Feb 27, 03:35 & 0.9993 & Wendelstein& $V  $ &  2$\times$600& 21.03$\pm$0.06\\
 2003 Feb 27, 03:58 & 1.0153 & Wendelstein& $B  $ &  2$\times$600& 21.33$\pm$0.11\\
 2003 Feb 27, 05:04 & 1.0541 & VLT/FORS1  & $R$   &  1$\times$120& 20.64$\pm$0.05\\
 2003 Feb 27, 05:05 & 1.0545 & JKT        & $R  $ & 1$\times$1200& 20.76$\pm$0.05\\
 2003 Feb 27, 05:04 & 1.0555 & NOT/ALFOSC & $V  $ &  1$\times$300& 21.11$\pm$0.02\\
 2003 Feb 27, 05:08 & 1.0580 & NTT/SuSI2  & $B  $ &  1$\times$300& 21.36$\pm$0.03\\
 2003 Feb 27, 05:10 & 1.0580 & VLT/FORS1  & $R$ + Wollaston & 8$\times$900 & \nodata \\
 2003 Feb 27, 05:13 & 1.0618 & NTT/SuSI2  & $B  $ &  1$\times$300& 21.37$\pm$0.03\\
 2003 Feb 27, 05:11 & 1.0621 & NOT/ALFOSC & $U  $ &  1$\times$600& 21.25$\pm$0.05\\
 2003 Feb 27, 05:18 & 1.0653 & NTT/SuSI2  & $B  $ &  1$\times$300& 21.48$\pm$0.04\\
 2003 Feb 27, 05:25 & 1.0701 & NTT/SuSI2  & $R  $ &  1$\times$300& 20.70$\pm$0.03\\
 2003 Feb 27, 05:29 & 1.0729 & NTT/SuSI2  & $R  $ &  1$\times$300& 20.68$\pm$0.03\\
 2003 Feb 27, 05:32 & 1.0750 & NTT/SuSI2  & $R  $ &  1$\times$300& 20.84$\pm$0.04\\
 2003 Feb 27, 06:14 & 1.1026 & VLT/FORS1  & $R$   &  1$\times$120& 20.79$\pm$0.05\\
 2003 Feb 27, 06:28 & 1.1156 & TNG        & $U  $ &  1$\times$600& 21.33$\pm$0.06\\
 2003 Feb 27, 08:10 & 1.1885 & NTT/SuSI2  & $B  $ &  3$\times$300& 21.73$\pm$0.03\\
 2003 Feb 27, 07:35 & 1.1910 & VLT/FORS1  & 600B  &  5$\times$900& \nodata \\
 2003 Feb 27, 08:27 & 1.1989 & NTT/SuSI2  & $R  $ &  3$\times$300& 20.99$\pm$0.02\\
 2003 Feb 27, 08:34 & 1.1997 & Tenagra II & $I  $ & 14$\times$300& 20.52$\pm$0.10\\
 2003 Feb 27, 09:50 & 1.2597 & UKIRT/UFTI & $K  $ & 18$\times$60 & 18.40$\pm$0.05\\
 2003 Feb 27, 12:55 & 1.3962 & UKIRT/UFTI & $K  $ & 36$\times$60 & 18.69$\pm$0.05\\
 2003 Feb 27, 23:49 & 1.8368 & NOT/ALFOSC & $V  $ &  1$\times$300& 22.50$\pm$0.07\\
 2003 Feb 27, 23:56 & 1.8434 & NOT/ALFOSC & $U  $ &  1$\times$600& 22.43$\pm$0.14\\
 2003 Feb 28, 03:14 & 1.9851 & Wendelstein& $R  $ &  2$\times$600& 21.89$\pm$0.13\\
 2003 Feb 28, 04:22 & 2.0246 & TNG        & $U  $ &  3$\times$600& 23.02$\pm$0.08\\
 2003 Feb 28, 11:47 & 2.3559 & UKIRT/UFTI & $K  $ & 54$\times$60 & 19.95$\pm$0.08\\
 2003 Mar 02, 04:58 & 4.0607 & VLT/FORS1  & $B  $ &  3$\times$600& 24.88$\pm$0.06\\
 2003 Mar 02, 05:32 & 4.0798 & VLT/FORS1  & $V  $ &  5$\times$180& 24.66$\pm$0.09\\
 2003 Mar 02, 05:52 & 4.0937 & VLT/FORS1  & $R  $ &  5$\times$180& 24.22$\pm$0.08\\
 2003 Mar 02, 06:13 & 4.1083 & VLT/FORS1  & $I  $ &  5$\times$180& 23.95$\pm$0.12\\
 2003 Mar 02, 06:53 & 4.1472 & VLT/ISAAC  & $K_s$ & 31$\times$10 & 21.32$\pm$0.21\\
 2003 Mar 03, 04:16 & 5.0753 &Gemini South& $V  $ &115$\times$60 & 24.64$\pm$0.10\\
 2003 Mar 05, 04:39 & 7.0500 & VLT/FORS1  & $R  $ & 10$\times$180& 25.26$\pm$0.15\\
 2003 Mar 13, 03:30 &$\!\!$15.0156 & VLT/FORS1  & $R  $ &  6$\times$600&$>$26.20 \\
\enddata
\tablenotetext{a}{mean observing epoch after the GRB trigger}
\end{deluxetable}

\clearpage

\begin{table*}
\centering
\caption[]{Line identifications\tablenotemark{a}}
\label{tab:lines}
\begin{small}
\begin{tabular}{l|l|cccc|cccc} 
\hline\\[-2mm]
     &                          & 
     \multicolumn{4}{c|}{Feb 26: FORS2 300~V} 
   & \multicolumn{4}{c}{Feb 27: FORS1 600~B}\\
Ion  & $f$ &
        $\lambda_{\rm obs}$ & $z_{\rm abs}$ & EW$_r$ & log $N$ &
        $\lambda_{\rm obs}$ & $z_{\rm abs}$ & EW$_r$ & log $N$ \\
      &  &  (\AA)            &               & (\AA)  & (cm$^{-2}$)  
         &  (\AA)            &               & (\AA)  & (cm$^{-2}$) \\
\hline\\[-3mm]
O I $\lambda$1302.17         & 0.049 & 3859 & 1.964 & 1.2$\pm$0.2 & 15.21$^{+
     0.07}_{-0.08}$ & 3856 & 1.962 & 0.9$\pm$0.2 & 15.09$^{+0.09}_{-0.11}$ \\
                       &       & 3892 & 1.989 & 2.2$\pm$0.2 & 15.48$^{+
     0.03}_{-0.05}$ & 3890 & 1.988 & 1.9$\pm$0.3 & 15.41$^{+0.07}_{-0.07}$\\[1mm]
C II $\lambda\lambda$1334.53,& 0.128,& 3952 & 1.961 & 1.1$\pm$0.2 & 14.74$^{+
     0.07}_{-0.09}$ & 3952 & 1.961 & 1.6$\pm$0.2 & 14.90$^{+0.05}_{-0.06}$\\
     \hspace*{1.2cm}    1335.71   & 0.115 & 3983 & 1.985 & 1.8$\pm$0.2 & 14.95$^{+
     0.05}_{-0.10}$ & 3985 & 1.986 & 2.1$\pm$0.3 & 15.02$^{+0.06}_{-0.07}$\\[1mm]
Si IV $\lambda$1393.75         & 0.528 & 4128 & 1.962 & 1.0$\pm$0.1 & 14.04$^{+
     0.04}_{-0.04}$ & 4126 & 1.961 & 1.0$\pm$0.2 & 14.04$^{+0.08}_{-0.09}$\\
                       &       & 4155 & 1.981 & 0.9$\pm$0.1 & 14.00$^{+
     0.04}_{-0.05}$ & 4153 & 1.980 & 1.1$\pm$0.2 & 14.08$^{+0.08}_{-0.08}$\\[1mm]
Si II $\lambda$1526.71         & 0.127 & 4522 & 1.962 & 0.9$\pm$0.2 & 14.54$^{+
     0.08}_{-0.11}$ & 4523 & 1.963 & 1.1$\pm$0.3 & 14.62$^{+0.09}_{-0.13}$\\
                       &       & 4558 & 1.986 & 1.4$\pm$0.1 & 14.73$^{+
     0.03}_{-0.03}$ & 4557 & 1.985 & 1.3$\pm$0.3 & 14.70$^{+0.09}_{-0.12}$\\[1mm]
C IV $\lambda\lambda$1548.20,& 0.191,& 4587 & 1.961 & 2.2$\pm$0.2 & 14.73$^{+
     0.04}_{-0.04}$ & 4585 & 1.964 & 1.1$\pm$0.2 & 14.43$^{+0.08}_{-0.08}$\\
     \hspace*{1.3cm}    1550.77   & 0.095 & 4629 & 1.988 & 1.0$\pm$0.1 & 14.39$^{+
     0.04}_{-0.04}$ & 4625 & 1.986 & $<$0.9& $<$14.35\\[1mm]
Fe II $\lambda$1608.45         & 0.058 & 4765 & 1.963 & 0.6$\pm$0.1 & 14.66$^{+
     0.06}_{-0.08}$ & 4764 & 1.962  & $<$0.8& $<$14.78\\
                           &        & 4802 & 1.986 & 0.8$\pm$0.1 & 14.78$^{+
     0.05}_{-0.06}$ & 4802 & 1.986  &$<$1.1 & $<$14.92\\[1mm]
Al II $\lambda$1670.79         & 1.880 & 4949 & 1.962 & 1.0$\pm$0.2 & 13.33$^{+
     0.08}_{-0.09}$ & 4952 & 1.964 & 0.7$\pm$0.1 & 13.18$^{+0.06}_{-0.07}$\\
                       &       & 4989 & 1.986 & 0.9$\pm$0.1 & 13.29$^{+
     0.04}_{-0.05}$ & 4989 & 1.986 & 1.2$\pm$0.2 & 13.41$^{+0.07}_{-0.08}$\\[1mm]
Mg II $\lambda$2796.35 & 0.612 & 5711 & 1.042 & 0.9$\pm$0.1 & 13.33$^{+0.04}_{-0.05}$ &--&--&--& --\\
                       &       & 8284 & 1.963 & 5.0$\pm$0.2 & 14.07$^{+0.02}_{-0.02}$ &--&--&--& --\\
                       &       & 8349 & 1.986 & 6.8$\pm$0.6 & 14.21$^{+0.03}_{-0.04}$ \\
Mg II $\lambda$2803.53 & 0.305 & 5726 & 1.042 & 0.5$\pm$0.1 & 13.37$^{+0.08}_{-0.09}$ &--&--&--& --\\
                       &       & 8305 & 1.962 & 3.5$\pm$0.3 & 14.22$^{+0.03}_{-0.04}$ &--&--&--& --\\
                       &       & 8370 & 1.985 & 5.9$\pm$0.4 & 14.44$^{+0.03}_{-0.03}$ \\[1mm]
\hline
\end{tabular}
\end{small}
\tablenotetext{a}{Symbols: oscillator strengths $f$ (Prochaska et al. 
2001), observer-frame wavelengths $\lambda_{\rm obs}$, 
corresponding absorption redshifts $z_{\rm abs}$, rest-frame
equivalent widths ($EW_r$) and deduced column densities ($N$[cm$^{-2}$]) of
the absorption lines detected in VLT spectra. Typical
errors in redshift are $\pm$0.001. Column densities were
calculated for the optically thin case. Note that the C~II and C~IV doublet
is not resolved in our spectra. The column densities for these ions refer to
the transition with the largest oscillator strength and neglect the other
component.}
\end{table*}

\end{document}